\documentclass[hyper]{JHEP3}
\pdfoutput=1
\usepackage{amsmath}
\usepackage{amsfonts}
\usepackage{amssymb}
\usepackage{amsthm}
\usepackage{mathtools}
\usepackage{graphicx}
\usepackage{ucs}
\usepackage{mathbbol}
\usepackage[utf8x]{inputenc}

\advance\hoffset0.25in

\makeatletter
\newdimen\itex@wd%
\newdimen\itex@dp%
\newdimen\itex@thd%
\def\itexspace#1#2#3{\itex@wd=#3em%
\itex@wd=0.1\itex@wd%
\itex@dp=#2ex%
\itex@dp=0.1\itex@dp%
\itex@thd=#1ex%
\itex@thd=0.1\itex@thd%
\advance\itex@thd\the\itex@dp%
\makebox[\the\itex@wd]{\rule[-\the\itex@dp]{0cm}{\the\itex@thd}}}
\makeatother

\makeatletter
\newif\if@sup
\newtoks\@sups
\def\append@sup#1{\edef\act{\noexpand\@sups={\the\@sups #1}}\act}%
\def\reset@sup{\@supfalse\@sups={}}%
\def\mk@scripts#1#2{\if #2/ \if@sup ^{\the\@sups}\fi \else%
  \ifx #1_ \if@sup ^{\the\@sups}\reset@sup \fi {}_{#2}%
  \else \append@sup#2 \@suptrue \fi%
  \expandafter\mk@scripts\fi}
\def\tensor#1#2{\reset@sup#1\mk@scripts#2_/}
\def\multiscripts#1#2#3{\reset@sup{}\mk@scripts#1_/#2%
  \reset@sup\mk@scripts#3_/}
\makeatother

\makeatletter
\newbox\slashbox \setbox\slashbox=\hbox{$/$}
\def\itex@pslash#1{\setbox\@tempboxa=\hbox{$#1$}
  \@tempdima=0.5\wd\slashbox \advance\@tempdima 0.5\wd\@tempboxa
  \copy\slashbox \kern-\@tempdima \box\@tempboxa}
\def\slash{\protect\itex@pslash}
\makeatother

\def\clap#1{\hbox to 0pt{\hss#1\hss}}

\def\mathrlap{\mathpalette\mathrlapinternal}
\def\mathclap{\mathpalette\mathclapinternal}

\def\mathrlapinternal#1#2{\rlap{$\mathsurround=0pt#1{#2}$}}
\def\mathclapinternal#1#2{\clap{$\mathsurround=0pt#1{#2}$}}

\let\oldroot\root
\def\root#1#2{\oldroot #1 \of{#2}}
\renewcommand{\sqrt}[2][]{\oldroot #1 \of{#2}}

\DeclareSymbolFont{symbolsC}{U}{txsyc}{m}{n}
\SetSymbolFont{symbolsC}{bold}{U}{txsyc}{bx}{n}
\DeclareFontSubstitution{U}{txsyc}{m}{n}

\DeclareSymbolFont{stmry}{U}{stmry}{m}{n}
\SetSymbolFont{stmry}{bold}{U}{stmry}{b}{n}

\makeatletter
\def\re@DeclareMathSymbol#1#2#3#4{%
    \let#1=\undefined
    \DeclareMathSymbol{#1}{#2}{#3}{#4}}
\re@DeclareMathSymbol{\neArrow}{\mathrel}{symbolsC}{116}
\re@DeclareMathSymbol{\neArr}{\mathrel}{symbolsC}{116}
\re@DeclareMathSymbol{\seArrow}{\mathrel}{symbolsC}{117}
\re@DeclareMathSymbol{\seArr}{\mathrel}{symbolsC}{117}
\re@DeclareMathSymbol{\nwArrow}{\mathrel}{symbolsC}{118}
\re@DeclareMathSymbol{\nwArr}{\mathrel}{symbolsC}{118}
\re@DeclareMathSymbol{\swArrow}{\mathrel}{symbolsC}{119}
\re@DeclareMathSymbol{\swArr}{\mathrel}{symbolsC}{119}
\re@DeclareMathSymbol{\nequiv}{\mathrel}{symbolsC}{46}
\re@DeclareMathSymbol{\Perp}{\mathrel}{symbolsC}{121}
\re@DeclareMathSymbol{\Vbar}{\mathrel}{symbolsC}{121}
\re@DeclareMathSymbol{\sslash}{\mathrel}{stmry}{12}
\re@DeclareMathSymbol{\invamp}{\mathrel}{symbolsC}{77}
\re@DeclareMathSymbol{\parr}{\mathrel}{symbolsC}{77}
\makeatother

\makeatletter
\DeclareRobustCommand\widecheck[1]{{\mathpalette\@widecheck{#1}}}
\def\@widecheck#1#2{%
    \setbox\z@\hbox{\m@th$#1#2$}%
    \setbox\tw@\hbox{\m@th$#1%
       \widehat{%
          \vrule\@width\z@\@height\ht\z@
          \vrule\@height\z@\@width\wd\z@}$}%
    \dp\tw@-\ht\z@
    \@tempdima\ht\z@ \advance\@tempdima2\ht\tw@ \divide\@tempdima\thr@@
    \setbox\tw@\hbox{%
       \raise\@tempdima\hbox{\scalebox{1}[-1]{\lower\@tempdima\box
\tw@}}}%
    {\ooalign{\box\tw@ \cr \box\z@}}}
\makeatother

\makeatletter
\def\udots{\mathinner{\mkern2mu\raise\p@\hbox{.}
\mkern2mu\raise4\p@\hbox{.}\mkern1mu
\raise7\p@\vbox{\kern7\p@\hbox{.}}\mkern1mu}}
\makeatother

\newcommand{\gt}{>}
\newcommand{\lt}{<}

\theoremstyle{plain}
\newtheorem{theorem}{Theorem}

\theoremstyle{definition}

\theoremstyle{remark}

\preprint{
UTTG-15-11 \\
TCC-017-11 \\
}
\title{Tinkertoys for the $D_N$ series}

\author{Oscar Chacaltana and Jacques Distler\\
      Theory Group and\\
      Texas Cosmology Center\\
      Department of Physics,\\
      University of Texas at Austin,\\
      Austin, TX 78712, USA \\
      {~}\\
      \email{oscarch@utexas.edu}\\
      \email{distler@golem.ph.utexas.edu}\\
      }

\date{June 27, 2011}

\abstract{
We describe a procedure for classifying 4D $\mathcal{N}=2$ superconformal theories of the type introduced by Davide Gaiotto. Any punctured curve, $C$, on which the 6D $(2,0)$ SCFT is compactified, may be decomposed into 3-punctured spheres, connected by cylinders. The 4D theories, which arise, can be characterized by listing the ``matter" theories corresponding to 3-punctured spheres, the simple gauge group factors, corresponding to cylinders, and the rules for connecting these ingredients together. Different pants decompositions of $C$ correspond to different S-duality frames for the same underlying family of 4D $\mathcal{N}=2$ SCFTs. In a previous work \cite{Chacaltana:2010ks}, we developed such a classification for the $A_{N-1}$ series of 6D $(2,0)$ theories. In the present paper, we extend this to the $D_N$ series. We outline the procedure for general $D_N$, and construct, in detail, the classification through $D_4$. We discuss the implications for S-duality in $Spin(8)$ and $Spin(7)$ gauge theory, and recover many of the dualities conjectured by Argyres and Wittig \cite{Argyres:2007tq}.
}

\begin{document}


\thispagestyle{empty}
\tableofcontents
\vfill
\newpage
\setcounter{page}{1}

\section{Introduction}\label{introduction}

Gaiotto duality \cite{Gaiotto:2009we, Gaiotto:2009hg,Gaiotto:2009gz,Nanopoulos:2010ga,Nanopoulos:2009uw,Chacaltana:2010ks,Tachikawa:2009rb,Tachikawa:2010vg,Benini:2010uu} identifies a large class of 4D $\mathcal{N}=2$ SCFTs with compactifications of the 6D $\mathcal{N}=(2,0)$ SCFT on a punctured Riemann surface, $C$. The moduli space, $\mathcal{M}_{g,n}$, parametrizes the family of exactly-marginal deformations of the SCFT. For every pants-decomposition of $C$, there is an $\mathcal{N}=2$ gauge-theoretic interpetation, in which each cylinder represents the  vector multiplets for some (simple) gauge group, and the 3-punctured spheres represent some sort of ``{}matter''{}, charged under the gauge groups of the attached cylinders. In particular, this construction identifies the boundaries of the moduli space, $\mathcal{M}_{g,n}$, with limits in which some, or all, of the gauge couplings become weak. Different degenerations correspond to different, S-dual, realizations of the same family of SCFTs.

Classifying the theories that arise, in this way, comes down to specifying (for a given 6D (2,0) theory) the 3-punctured spheres, the gauge groups associated with the cylinders that connect them, and the rules for gluing these ingredients together. Arbitrarily complicated 4D $\mathcal{N}=2$ SCFTs can be constructed, in ``{}tinkertoy''{} fashion, by connecting together these basic ingredients.

For a given (2,0) theory, this is a finite task. In our previous paper \cite{Chacaltana:2010ks}, we carried out this program for theories that are obtained from a compactification of the (2,0) theories of type $A_{N-1}$. In so-doing, we identified a multitude of new interacting, non-Lagrangian SCFTs (generalizing \cite{Minahan:1996cj}), corresponding to compactifications of the $A_{N-1}$ theory on certain 3-punctured spheres. Their appearance, in the context of Gaiotto duality, is a vast generalization of the classic examples of non-Lagrangian SCFTs appearing in the S-dual description of more-familiar $\mathcal{N}=2$ gauge theories, discovered by Argyres and Seiberg \cite{Argyres:2007cn}.

While Gaiotto'{}s original arguments relied on the realization of the 6D theory as the low-energy theory of $N$ M5-branes, which necessarily implied working with a 6D theory of $A_{N-1}$ type, the idea can be straightforwardly generalized to the case of $N$ M5 branes in the presence of an orientifold, whose low-energy limit is the 6D theory of type $D_N$. (There is, by contrast, no known realization of the 6D theories of type $E$ as a low-energy theory of M5 branes.) The class of 4D SCFTs arising from the compactification of the $D_N$ 6D theories on Riemann surfaces has been considerably less studied \cite{Tachikawa:2009rb,Tachikawa:2010vg,Benini:2010uu} than its $A_{N-1}$ analogue.

As for the $A_{N-1}$ theories, the Seiberg-Witten curve of 4D theories arising from the $D_N$ theories can be written in Gaiotto'{}s form, as a polynomial equation in the Seiberg-Witten differential (a 1-form on $T^*C$), whose coefficients are (the pullbacks of) differentials on $C$. The differentials descend from protected operators of the 6D theory, and so their degrees are equal to the exponents of $Spin(2N)$.

Just as Gaiotto used the well-known $SU(n)$ linear quivers to test his arguments for the $A_{N-1}$ theory, Tachikawa \cite{Tachikawa:2009rb,Tachikawa:2010vg} studied the SO-Sp linear quivers \cite{Landsteiner:1997vd, Brandhuber:1997cc} to find the pole structure and flavour symmetry group for punctures in the $D_N$ theory, and discovered a few examples of S-duality. Unfortunately, the SO-Sp linear quivers, that arise from the orientifold construction, live in a theory slightly larger than the one we are interested in. The $A_{N-1}$, $D_N$ and $E_6$ theories have a $\mathbb{Z}_2$ outer-automorphism group (which gets enhanced to $S_3$ in the case of $D_4$), and we can consider compactifications of the (2,0) theory, where going around a homologically-nontrivial cycle on $C$ (circumnavigating a handle, or circling a puncture) is accompanied by an outer-automorphism twist.

A proper discussion of the incorporation of outer-automorphism twists should treat the $A_{N-1}$, $D_N$ and $E_6$ (2,0) theories in tandem, as all of these Dynkin diagrams have a $\mathbb{Z}_2$ outer automorphism. We will leave that discussion to future work\footnote{See \cite{Chacaltana:2012ch} for a treatment of the $\mathbb{Z}_2$-twisted $A_{2N-1}$ series.}. Instead, in this paper, we will study the compactifications of the $D_N$ theory, \emph{without} outer-automorphism twists, and develop a classification precisely analogous to the one we developed for the $A_{N-1}$ theory (also without outer automorphism twists). Nonetheless, at a crucial point, we will have recourse to Tachikawa'{}s linear quiver tail analysis which, strictly speaking, embeds the $D_N$ theories without outer automorphism twists in the larger class of $D_N$ theories which \emph{do} include outer automorphism twists.

The analysis in the $D_N$ case introduces several new complications, not seen in the $A_{N-1}$ case. In the $A_{N-1}$ theory, each puncture corresponded to a choice of partition of $N$ (equivalently, to an $N$-box Young diagram, or a nilpotent orbit in the complexified Lie algebra, $\mathfrak{sl}(N)$). The chosen partition determined the ``{}flavour symmetry''{} group (essentially, the isometry group of the Higgs branch) associated to a given puncture. At the same time, it (or, more accurately, its \emph{transpose}) determined the singular behaviour of the Hitchin system at the puncture which, in turn, gave the geometry of the Coulomb branch.

In the present case, that relationship is more complicated. As in the $A_{N-1}$ case, the flavour symmetry group (geometry of the Higgs branch) is determined by a ``{}D-partition''{} of 2N. Such partitions also label nilpotent orbits in $\mathfrak{so}(2N)$. However, only for a subset of these, the ``{}special''{} D-partitions \cite{Collingwood1993}, is the behaviour of the Hitchin system at the puncture given by (the ``Spaltenstein dual'') nilpotent orbit.

The Coulomb branch of the theory comprises the degrees of freedom associated to a set of meromorphic $k$-differentials on the Riemann surface which are allowed to have poles of certain orders (determined by the choice of partition) at the punctures. A new feature, of the $D_N$ case, is that the coefficients of the leading poles of these differentials obey certain polynomial constraints. The ``{}true''{} Coulomb branch is obtained, after imposing the constraints.

These constraints were derived by Tachikawa \cite{Tachikawa:2009rb}, by considerations involving linear quiver tails. We will present a slightly different, more intrinsic, viewpoint on the origin of these constraints. For the special partitions, we will see that the constraints pop out naturally from requiring that the Higgs field have a simple pole with residue lying in the Spaltenstein-dual nilpotent orbit. For the non-special partitions, we will content ourselves with determining the pole structure of the $k$-differentials at the puncture, and the associated constraints, using the linear quiver tail analysis. We refer to \cite{Chacaltana:2012zy} for a full discussion of the boundary condition of the Hitchin system for non-special punctures.

A further peculiar feature of the non-special punctures is that the global symmetry group of the puncture contains ${Sp(l)}_k$ factors, with $k$ odd. This level for the current algebra is that which would be induced by an odd number of half-hypermultiplets in the fundamental $2l$-dimensional representation. In other words, this symmetry is subject to Witten'{}s global anomaly \cite{Witten:1982fp} and (in the absence of additional matter) could not be consistently gauged.

Even after having dealt with these new complexities, simply enumerating the \emph{results} in the $D_N$ case is considerably more tedious than it was in the $A_{N-1}$ case. The number of fixtures (3-punctured spheres), and the number of cylinders that connect them, proliferate much more rapidly with $N$.

We will restrict ourselves to presenting a complete catalogue only for $D_4$. As a measure of the complexity, there are 99 3-punctured spheres for $D_4$; we will list all of those. There are 785 4-punctured spheres ---{} theories with a single gauge group factor ---{} it would be prohibitive to list all of those.

Nevertheless, $D_4$ is an interesting case to study. As we said before, the outer automorphism group is enhanced to $S_3$. This group is a symmetry of the $D_4$ (2,0) theory, and so acts on the set of punctures/fixtures/cylinders, which are naturally organized into multiplets, permuted by the outer automorphisms. As already mentioned, we will \emph{not} consider the inclusion of outer-automorphism \emph{twists}.

For the $D_5$ and $D_6$ theories, we will present tables of the regular punctures and their properties, but will refrain from presenting a complete catalogue of fixtures and cylinders.

As in the $A_{N-1}$ series, we discover several new interacting SCFTs ---{} non-Lagrangian fixed points of the renormalization group ---{} and we realize a number of S-dualities predicted by Argyres and Wittig \cite{Argyres:2007tq}. We also provide formul\ae\ for the conformal-anomaly central charges $a, c$, and explain how to compute the flavour current-algebra charges $k$, for interacting SCFTs.

\section{{The $D_N$ Series}}\label{D-NSeries}

Much of the construction is well-reviewed in previous works \cite{Gaiotto:2009we, Gaiotto:2009hg, Tachikawa:2009rb, Tachikawa:2010vg, Nanopoulos:2009uw, Nanopoulos:2010ga, Nanopoulos:2010bv, Benini:2010uu}, so we will be somewhat brief, concentrating on the novelties which arise in the $D_N$ case. We consider a 6D (2,0) theory compactified on a Riemann surface $C$ of genus $g$ with $n$ punctures (complex codimension-1 defect operators) located at points $y_i\in C,\,i=1,\dots,n$.

In the $A_{N-1}$ case, the Seiberg-Witten curve, $\Sigma\subset T^*C$ of the 4D low-energy theory is given by
\begin{equation}
0=\lambda^N  + (-1)^{N} \sum_{k=2}^{N} \lambda^{N-k} \phi_k(y),
\label{ANSW}\end{equation}
where $\lambda$ is the Seiberg-Witten differential, and the $\phi_k(y)$ are $k$-differentials on $C$ (pulled back to $T^*C$). The $\phi_k$ are allowed to have poles of various orders at the $y_i$.

The theory possesses a set of relevant operators, whose vacuum expectation values parametrize the Coulomb branch of the theory. At a generic point on the Coulomb branch, the theory is infrared-free; at the origin, it is superconformal. The tangent space at the origin of the Coulomb branch is a graded vector space,

\begin{equation}
V=\bigoplus_{k=2}^N V_k.
\label{linspace}\end{equation}
where $V_k = H^0\Bigl(C, K^k\left(\sum_{i=1}^n p_k^{(i)}y_i\right)\Bigr)$ is the vector space of meromorphic of $k$-differentials, $\phi_k$, with poles of order at most $p^{i}_k$ at the punctures $y_i$.

As we vary the gauge couplings, the graded vector spaces, $V$, fit together to form the fibers of a graded vector bundle over the moduli space, $\mathcal{M}_{g,n}$, of marginal-deformations. Our main guiding principle is that this vector bundle should extend to the boundary of $\mathcal{M}_{g,n}$. What naturally extends, over $\overline{\mathcal{M}}_{g,n}$, are the virtual bundles whose fibers are

$$
H^0\Bigl(C, K^k\bigl(\sum_{i=1}^n p_k^{(i)}y_i\bigr)\Bigr)\ominus H^1\Bigl(C, K^k\bigl(\sum_{i=1}^n p_k^{(i)}y_i\bigr)\Bigr)\quad .
$$

We will arrange for the $H^1$s to vanish, so that the virtual bundle is an honest bundle, which extends to the boundary. At the boundary, the Coulomb branch has components associated to the irreducible components of $C$ and components associated to the gauge groups on the degenerating cylinders.

For the $D_N$ series of $(2,0)$ theories, the story is superficially similar. The Seiberg-Witten curve takes the form

\begin{equation}
0=\lambda^{2N} + \sum_{k=1}^{N-1} \lambda^{2(N-k)} \phi_{2k}(y) +\tilde{\phi}^2(y)
\label{DNSW}\end{equation}
Again, the $\phi_{2k}$ and $\tilde{\phi}$ are meromorphic differentials on $C$, with poles of up to the prescribed orders at the punctures. ($\tilde{\phi}$ is the Pfaffian, i.e., an $N$-differential.)

However, there are some crucial differences between the $A_{N-1}$ and $D_N$ theories. While in the $A_{N-1}$ case, the coefficients in the Seiberg-Witten equation \eqref{ANSW} were just linear functions of the Coulomb branch \eqref{linspace}, in the $D_N$ case, the coefficients in Seiberg-Witten equation \eqref{DNSW} are, in general, polynomial expressions when expressed in terms of the natural linear coordinates at the origin of the Coulomb branch. We see that, already, in the fact that the Seiberg-Witten equation depends quadratically on $\tilde{\phi}$. But there are further polynomial constraints on the coefficients in the $\phi_{2k}$, which need to be solved before one sees the natural linear structure.

While the constraints are polynomial, they are always \emph{linear} in (at least) \emph{one} of the variables. Moreover, they are of homogeneous degree in the aforementioned grading. So the space of solutions of the constraints is always smooth at the origin of the Coulomb branch, and hence the tangent space at the origin has the desired structure of a graded vector space.

The other complication in the $D_N$ theories is that, whereas the differentials in the $D_N$ theory have degrees $2,4,6,\dots,2(N-1); N$, the Coulomb branch has components in other degrees. For instance, in $D_4$, there is a component of degree 3, in addition to the ``{}expected''{} components of degrees $2,4,6$. In general, the Coulomb branch takes the form

\begin{displaymath}
E\subset V
\end{displaymath}
where

\begin{displaymath}
V = \bigoplus_{k=1}^{N-1} H^0\left(C,K^{2k}\bigl(\sum_{i=1}^n p^{(k)}_i y_i\bigr)\right)\quad \oplus\quad \bigoplus_{k=3}^{N-1} W_k\quad \oplus\quad H^0\left(C,K^{N}\bigl(\sum_{i=1}^n \tilde{p}_i y_i\bigr)\right)
\end{displaymath}
Here the $W_k$ are vector spaces of degree $k$ and $E$ is the subvariety satifying the collection of polynomial constraints (linear in at least one variable, and of homogeneous degree).

If we denote the coefficient of $l^{\text{th}}$-order pole of $\phi_k$, at one of the punctures, by $c^{(k)}_l$, the constraints can roughly be divided into

\begin{itemize}%
\item polynomials (of homogeneous degree in both $k$ and $l$) in the $c^{(k)}_l$
\item polynomials (again, of appropriately homogeneous degree) involving both the $c^{(k)}_l$ and a basis $a^{(k)}$ for the vector spaces, $W_k$

\end{itemize}
In the case of $D_4$, there is just $W_3$, and $dim(W_3) = n_o$, the number of punctures, on $C$, corresponding to a particular special D-partition. At each such puncture, there is a constraint $c^{(6)}_4 = {\left(a^{(3)}\right)}^2$, which says that the coefficient of the leading singularity of $\phi_6$ is the square of a gauge-invariant quantity, $a^{(3)}$, of scaling dimension three. 

\subsection{{Punctures and the Spaltenstein Map}}\label{spaltenstein}

In the $A_{N-1}$ series, punctures are labeled by partitions of $N$. To each such partition, $[h_1,h_2,\dots h_p]$, with

\begin{displaymath}
\begin{gathered}
  h_1\geq h_2\geq\dots\geq h_p,\\
  \sum_{i=1}^p h_i = N,
\end{gathered}
\end{displaymath}
we associated a Young diagram, whose $i^{\text{th}}$ column has height $h_i$. The corresponding flavour symmetry group is

\begin{equation}
G= S\left(\prod_h U(n^{(h)})\right),
\label{AflavourSym}\end{equation}
where $n^{(h)}$ is the number of columns of height $h$. We call the partition $[h_1,h_2,\dots h_p]$, which labels the puncture, the \emph{Nahm partition}\footnote{This nomenclature is justified in \cite{Chacaltana:2012zy}.} for the puncture. We emphasize that when we represent a Nahm partition by a Young diagram, we will always take its parts to be the \emph{column-heights} of the Young diagram.

Of course, a Young diagram with column-heights $[h_1,h_2,\dots, h_p]$ determines a second partition of $N$, given by the row-lengths, $[r_1, r_2,\dots, r_q]$. The two partitions are said to be \emph{transposes} of each other.

This second partition determines a nilpotent orbit \cite{Collingwood1993}, $\mathcal{O}_{[r_1, r_2,\dots, r_q]}\subset \mathfrak{sl}(N)$, which determines the pole structure of the $\phi_k(y)$ at the puncture. Specifically, the Higgs field, $\varphi$, of the Hitchin system on $C$
has a simple pole at the puncture, with residue $X$ lying on the nilpotent orbit $\mathcal{O}_{[r_1, r_2,\dots, r_q]}$  \cite{Gukov:2006jk, Gaiotto:2009hg, Nanopoulos:2009uw},
\begin{equation}
\varphi(y)=\frac{X}{y} + \text{generic},
\label{higgsfield}\end{equation}
where $y$ is a local coordinate on $C$ such that the puncture is at $y=0$, and we allow for a generic element (a regular function of $y$) in $\mathfrak{sl}(N)$.

We call the partition $[r_1, r_2,\dots, r_q]$, which determines the boundary condition for the Hitchin system, the \emph{Hitchin partition} of the puncture. When we want to represent a Hitchin partition by a Young diagram, we will always take its parts to be the \emph{row-lengths}. For a puncture in the $A_{N-1}$ series, the Hitchin partition is simply the transpose of the Nahm partition, and both are represented by the same Young diagram\footnote{Let us remark that \emph{any} partition of $N$ corresponds to a nilpotent orbit of $\mathfrak{sl}(N)$ \cite{Collingwood1993}, so, in particular, the Nahm partition of a puncture also corresponds to a nilpotent orbit. However, it is the nilpotent orbit associated to the Hitchin partition that is relevant to the Hitchin system boundary condition \eqref{higgsfield}. Also, the fact that both the Nahm and the Hitchin partitions can be represented by the same Young diagram is a peculiarity of the $A_{N-1}$ series, and does not extend to the $D_N$ series, as we will soon see.}.

There is a fairly simple algorithm for choosing the nilpotent representative $X$ in terms of the Hitchin partition:

\begin{itemize}%
\item Let $X$ be a block-diagonal matrix, where the $i^{\text{th}}$ block is $r_i\times r_i$.
\item Within each block, let $X$ be strictly upper-triangular.

\end{itemize}
The characteristic equation for $\varphi$,

\begin{equation}
\det(\varphi(y) - q \mathbb{1}) = (-q)^N + \sum_{k=2}^N q^{N-k} \phi_k(y),
\end{equation}
which yields the Seiberg-Witten equation \eqref{ANSW}, determines also the allowed pole orders of the $\phi_k$. The resulting list of pole orders is easily expressed in terms of the Young diagram:

\begin{itemize}%
\item Starting with 0 in the first box, number the boxes in the first row with successive positive integers.
\item When you get to the end of a row, repeat that integer as the number assigned to the first box of the succeeding row. Continue numbering the boxes of that row with successive integers.
\item The integers inscribed in boxes $2,\dots, N$ are, respectively, the pole orders of $\phi_2,\dots, \phi_N$.

\end{itemize}
For the $D_N$ series, punctures are labeled by partitions of $2N$. However, not all partitions are allowed.

\begin{itemize}%
\item Even integers must occur with even multiplicity.
\item When all the integers in the partition are even (such a partition is called ``{}very even''{}), we get \emph{two} punctures. Such partitions only occur for $N$ even. These two punctures are exchanged by the $\mathbb{Z}_2$ outer automorphism of $D_N$ which exchanges the two spinor representations. We will colour the corresponding Young diagrams red and blue, to distinguish them.

\end{itemize}
Such a partition is called a ``{}D-partition of $2N$.''{} So, we say that a puncture in the $D_N$ series is labeled uniquely by a \emph{Nahm} D-partition of $2N$, except in the case of a ``very-even'' Nahm D-partition, which corresponds to \emph{two} punctures, and so requires an additional label to distinguish them. As before, if we wish to represent a Nahm D-partition by a Young diagram, its parts give the column-heights of the Young diagram. For very-even Nahm D-partitions, we will colour the Young diagram in red/blue, to distinguish the two punctures labeled by it.

On the other hand, it is known \cite{Collingwood1993} that a D-partition of $2N$ labels nilpotent orbits in $\mathfrak{so}(2N)$, except in the case of a very-even D-partition, which corresponds to two nilpotent orbits, and, again, an additional label is needed to distinguish them. So, if we wish, punctures in the $D_N$ series are labeled uniquely by nilpotent orbits in $\mathfrak{so}(2N)$, which we call \emph{Nahm} nilpotent orbits in $\mathfrak{so}(2N)$.


From the Young diagram corresponding to the Nahm D-partition of a puncture, we reconstruct the flavour symmetry group, associated to the puncture,

\begin{equation}
G = \prod_{\mathclap{h\, \text{odd}}} Spin\left(n^{(h)}\right)\times
\prod_{\mathclap{h\, \text{even}}} Sp\left(\tfrac{n^{(h)}}{2}\right).
\label{GflavourDef}\end{equation}
From this, the necessity of the the rule that $n^{(h)}$ be even, for even $h$, is obvious. The origin of the additional rule (which arises for $N$ even) ---{} that ``{}very even''{} D-partitions occur twice ---{} has a more subtle origin.

For $N$ odd, the irreducible spinor representation of $D_N$ is complex, and the right-handed spinor representation is the complex-conjugate of the left-handed one. So a ``{}hypermultiplet in the spinor''{} contains fields transforming as spinors of both chiralities.

For $N$ even, the irreducible spinor representation is real ($N=4l$) or pseudoreal ($N=4l+2$), and the left- and right-handed spinor representations are inequivalent. So a ``{}hypermultiplet in the left-handed spinor representation''{} is \emph{different} from a ``{}hypermultiplet in the right-handed spinor representation.''{} When we discuss fixtures, we will need to keep track of this distinction. Exchanging ``{}red''{} and ``{}blue''{} punctures will exchange the roles of left- and right-handed spinors.

Understanding the singularities of the $\phi_k$ at the puncture is somewhat more involved than in the $A_{N-1}$ case.

As in the $A_{N-1}$ case, we might expect to associate a D-partition of $2N$ (or, equivalently, a nilpotent orbit in $\mathfrak{so}(2N)$) to the \emph{rows} of the Nahm Young diagram. Unfortunately, when the \emph{columns} of a $2N$-box Young diagram form a D-partition, the \emph{rows} typically do not. In other words, the transpose does not map D-partitions to D-partitions. Nevertheless, there is a simple modification of the transpose map, called the ``{}Spaltenstein map''{} which \emph{does} map D-partitions to D-partitions.

This procedure may be described as (row) ``{}D-collapse''{}:

\begin{itemize}%
\item Given a Nahm Young diagram (that is, one whose column-heights form a D-partition), take the longest even row, which occurs with odd multiplicity (if the multiplicity is greater than 1, take the \emph{last} row of that length), and remove the last box. Place the box at the end of the next available row, such that the result is a Young diagram.
\item Repeat the process with next longest even row, which occurs with odd multiplicity.
\item This process eventually terminates, and the result is a ``{}corrected''{} Young diagram (which we call \emph{Hitchin} Young diagram), whose row-lengths form a \emph{Hitchin} D-partition.

\end{itemize}
Conversely, starting with a Hitchin Young diagram (i.e., whose rows form a D-partition)
, we can define a process of \emph{column D-collapse}, which yields a Nahm Young diagram (whose columns form a D-partition).

In the $A_{N-1}$ case, the Spaltenstein map was given by the transpose.
In the $D_{N}$ case, the Spaltenstein map is defined as the composition of the transpose with the 
D-collapse. Unfortunately, unlike the transpose, the Spaltenstein map is \emph{not} an involution of the set of D-partitions; in general, it is neither 1-1 nor onto. The set of partitions in the image of the Spaltenstein map are called ``{}special''{}, and the Spaltenstein map, restricted to the special partitions, \emph{is} an involution.

More formally, let $s$ be the Spaltenstein map, and let $p$ be a D-partition. $p$ is called ``{}special''{} if $s^2(p)=p$. In the $A_{N-1}$ case, all partitions were special (${(p^t)}^t = p$). That is not the case for $D_N$. Instead, we have the theorem 

\begin{theorem}{(\cite{Collingwood1993} Corollary 6.36 and Proposition 6.3.7)}\label{theorem}
\begin{enumerate}
\item For any D-partition, $p$, $s(p)$ is a special D-partition.
\item A D-partition, $p$, is special, if and only if $p^t$ is a C-partition. (A \emph{C-partition} of $2N$ is a partition with the property that \emph{odd} integers occur with even multiplicity.)
\end{enumerate}
\end{theorem}
The Hitchin system boundary conditions for punctures labeled by \emph{special} D-partitions are determined as in the $A_{N-1}$ case. Let $f$ be the Nahm D-partition, and let $o=s(f)$ be the Hitchin nilpotent orbit, that is, the image of $f$ under the Spaltenstein map. If $f$ is special (which was \emph{always} the case for $A_{N-1}$), then the Higgs field $\varphi(y)$ has a simple pole, with residue $X\in o$, exactly as in \eqref{higgsfield}, except that the generic element now lives in $\mathfrak{so}(2N)$. Under the obvious embedding $\mathfrak{so}(2N)\hookrightarrow \mathfrak{sl}(2N)$, the characteristic equation

\begin{equation}
\det(\varphi(y) - q \mathbb{1}) = q^{2N} + \sum_{k=1}^{N-1} q^{2(N-k)} \phi_{2k}(y)\, +\, {(\tilde{\phi}(y))}^2,
\label{DNSWHiggs}\end{equation}
which reproduces the Seiberg-Witten curve \eqref{DNSW}, yields the pole orders of the $k$-differentials\footnote{Since it will be important for us to keep track of the sign of $\tilde{\phi}(y)$, it is best to compute it separately. In the antisymmetric basis of $\mathfrak{so}(2N)$, we have
$$
\tilde{\phi}(y)=\frac{1}{2^N N!}\sum_{\pi\in S_{2N}}\text{sgn}(\pi)\prod_{i=1}^{N}\left(\varphi(y)\right)_{\pi(2i-1),\pi(2i)},
$$
where $S_{2n}$ are all permutations of $\{1,\dots,2N\}$.}. These can be read off from the Hitchin Young diagram, just as if it were a Young diagram for $A_{2N-1}$. (See the rule above.) Because $\varphi(y)$ lies in the $\mathfrak{so}(2N)$ subalgebra, the $\phi_k$ vanish for odd $k$, and $\phi_{2N}(y) = {(\tilde{\phi}(y))}^2$. That, however, does not quite exhaust the constraints on the polar parts of the $k$-differentials, which follow from restricting to $\mathfrak{so}(2N)\subset \mathfrak{sl}(2N)$. There are additional polynomial constraints among the coefficients of the leading-order poles of the various $k$-differentials.

These additional constraints were previously found by Tachikawa \cite{Tachikawa:2009rb} by applying the restrictions, imposed by M-theory orientifolds \cite{Hori:1998iv}, to SO-Sp linear quiver tails. As already mentioned, the SO-Sp quivers naturally live in the larger theory, with outer-automorphism twists. From our present perspective it is better to think of the constraints as coming directly from putting the polar part of $\varphi(y)$ in a special nilpotent orbit of $\mathfrak{so}(2N)$. (For our explicit conventions on nilpotent orbits in $\mathfrak{so}(2N)$, see  \S\ref{appendix}.)

As a simple example, consider the minimal\footnote{This puncture is ``minimal'' in the sense that its Spaltenstein dual is the smallest non-trivial (Hitchin) nilpotent orbit. This nomenclature agrees with that of the existing $D_N$-series literature (e.g., \cite{Tachikawa:2009rb,Tachikawa:2010vg}).} $D_3$ puncture, which has special Nahm Young diagram $\includegraphics[width=17pt]{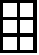}$. To find its pole structure, we put the polar part of the Higgs field in the nilpotent orbit of the Spaltenstein dual, corresponding to the Hitchin Young diagram
$$\includegraphics[width=17pt]{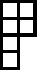}$$
We write $\varphi(y)=\frac{X}{y}+M$, where $X=X^{-}_{1,2}$ is the canonical nilpotent element in this orbit (see \S\ref{appendix} for our conventions), and $M$ is a generic matrix in $\mathfrak{so}(2N)$, of the form \eqref{so2Nalg}. The differentials are thus of the form
\begin{equation}
\phi_2 =\frac{2a}{y}+\dots,\qquad
\phi_4 =\frac{a^2}{y^2}+\dots,\qquad
\tilde{\phi} =\frac{b}{y}+\dots
\end{equation}
Hence, the pole structure is $\{1,2;1\}$, with a constraint $c^{(4)}_{2}=\frac{1}{4}\left(c^{(2)}_{1}\right)^2$. This pole structure and constraint were computed in \cite{Tachikawa:2009rb} from the SO-Sp linear quiver tail for this puncture.

That takes care of the \emph{special} punctures, that is, those labeled by a special Nahm D-partition. What about \emph{non-special} punctures, i.e., the ones labeled by non-special Nahm D-partitions? Here the situation is a bit more awkward. The Spaltenstein map is not an involution. When applied to a non-special partition, the image is a special partition, and there are several Nahm partitions that map to the same (special) Hitchin D-partition. To faithfully preserve the information of the original Nahm partition, one needs to supplement the Spaltenstein map by some additional discrete data. We leave the details of this problem to \cite{Chacaltana:2012zy}. The effect on the pole structure of the $k$-differentials, however, is easy to find (say, from the linear quiver tail analysis), and amounts to the following. Given a non-special Nahm D-partition, $f$, $f_s= s^2(f)$ is a special Nahm D-partition. The pole structure of the $\phi_k(y)$ for the non-special puncture $f$ is precisely that one would find for the special puncture $f_s$. However, $f_s$ has a series of constraints of the form $c^{(2k)}_{2l} = {\left(a^{(k)}\right)}^2$ on the leading pole coefficients. For the non-special puncture, $f$, some (or all) of these constraints are \emph{relaxed}.

To see which constraint(s) are relaxed, notice that the Nahm Young diagram for $f_s$ can be obtained from that for $f$ by a process of (row) C-collapse. That is, we remove the last box from a row of odd length (which occurred with odd multiplicity) and place it lower-down on the Young diagram. The box we removed was an odd-numbered box (call it $2k+1$). By removing it, an even-numbered box (box $2k$) becomes the last box in that row. The puncture, $f_s$, had a constraint of the form $c^{(2k)}_{2l} = {\left(a^{(k)}\right)}^2$. For each ${(2k)}^{\text{th}}$ box, thus exposed, we relax the corresponding constraint of $f_s$.

For instance, for $D_4$, there is just one non-special puncture and, correspondingly, just one constraint that gets relaxed. Plenty of other examples can be seen in the tables of Sec.~\ref{regular}.

Finally, let us elaborate on our conventions for ``very even" punctures. When $N$ is even, the Pfaffian, $\tilde{\phi}$ has the same degree as $\phi_N$. The outer-automorphism of $D_N$, which exchanges the roles of the two spinor representations, takes

\begin{equation}
\begin{aligned}
  \tilde{\phi}&\mapsto - \tilde{\phi}\\
  \phi_{2k}&\mapsto \phi_{2k},\qquad k=1,\dots, N-1
\label{Z2actionphik}\end{aligned}
\end{equation}
For most punctures, the contraints are such that there is a \emph{unique} Coulomb branch parameter (the coefficient $c^{(2k)}_l$ of the highest-order pole of one of the $\phi_{2k}$) which appears linearly. We can then take $c^{(2k)}_l$ to be the variable eliminated by the constraint, so for the purpose of counting the graded dimension of the Coulomb branch, it is as if we simply reduced the allowed pole-order, $p_{2k}$, for $\phi_{2k}$ by 1.

Certain red/blue punctures are an exception. At these punctures, both $\tilde{\phi}$ and $\phi_N$ are allowed to have poles of some order (say, $l$), but a \emph{linear combination} of the coefficients, $c^{(N)}_l\pm 2\tilde{c}_l$, is the variable that appears linearly in the associated constraints, which are of the form
\begin{equation}
c^{(N)}_l\pm 2\tilde{c}_l=\dots,
\label{redbluelinearconstraint}
\end{equation}
where the ellipsis stands for additional terms. The signs above may correspond to red or blue, depending on the case. At any rate, because of \eqref{Z2actionphik}, the full sets of constraints for red and blue punctures with the same Nahm D-partition are related by $\tilde{c}_l\to -\tilde{c}_l$.

As an example, let us look at the punctures with Nahm Young diagrams $\includegraphics[width=17pt]{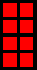}$ and  $\includegraphics[width=17pt]{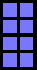}$, which are the same as their Hitchin Young diagrams\footnote{We refrain from arguing whether a Nahm red/blue D-partition should map to a Hitchin red or blue D-partition. While \cite{Collingwood1993} contends that it is natural to define the $D_{2k}$ Spaltenstein map to take $\{\text{red},\text{blue}\}\mapsto\{\text{red},\text{blue}\}$ for $k$ even, and $\{\text{red},\text{blue}\}\mapsto\{\text{blue},\text{red}\}$ for $k$ odd, it is possible that the physical map Nahm$\to$Hitchin be the Spaltenstein map composed with the $\mathbb{Z}_2$-action that exchanges red and blue. So, in this paper, we simply assume that a Nahm red (blue) puncture maps to a Hitchin red (blue) puncture.}. The canonical nilpotent elements (see \S\ref{appendix}) are $X^{(r)}=X^{-}_{1,2}+X^{-}_{3,4}$ and $X^{(b)}=X^{-}_{1,2}+X^{+}_{3,4}$, respectively. After writing $\varphi(y)=\frac{X^{\mathrlap{\text{(r/b)}}}}{y}\,\, +M$ for the Higgs field, with $M$ a generic $\mathfrak{so}(2N)$ matrix, we find for the differentials,
\begin{equation}
\begin{aligned}
\phi_2 & =\frac{2a}{y}+\dots\\
\phi_4 & =\frac{a^2\mp 2b}{y^2}+\dots\\
\phi_6 & =\frac{\mp 2ab}{y^3}+\dots\\
\tilde{\phi} & =\frac{b}{y^2}+\dots
\end{aligned}
\end{equation}
with the top sign for the red and the lower sign for the blue puncture. So the pole structure for these punctures is $\{1,2,3;2\}$, with constraints $c^{(4)}_{2}\pm 2\tilde{c}_2=\frac{1}{4}(c^{(2)}_{1})^2$ and $c^{(6)}_{3}=\mp \tilde{c}_{2} c^{(2)}_{1}$. The $\mathbb{Z}_2$ outer automorphism acts as $b\mapsto -b$, and it exchanges the red and blue constraints.

In the presence of red/blue punctures with constraints of the form \eqref{redbluelinearconstraint}, a little extra care must be taken in computing the graded Coulomb branch dimensions. Too large an excess, of one or the other, over-constrains the differentials and would lead to a difference between the virtual and actual dimension of the Coulomb branch. The dimension of the degree-$N$ component is
\begin{equation}
\operatorname{dim}(V_N) = d_N + \tilde{d} - n_r -n_b,
\end{equation}
where $d_N$ and $\tilde{d}$ are the dimensions we would obtain from applying Riemann-Roch (suitably-adjusted for the other constraints) to $\phi_N$ and $\tilde{\phi}$, and $n_{r,b}$ are the number of constraints of the form $c^{(N)}\pm 2\tilde{c}_l=\dots$ for red and blue punctures, respectively. In order that the constraints not be over-determined, it suffices to ensure that either
\begin{equation}
d_N - n_r \geq 0,\quad \tilde{d} - n_b \geq 0
\end{equation}
or
\begin{equation}
d_N - n_b \geq 0,\quad\tilde{d} - n_r \geq 0
\end{equation}
holds. Either condition is sufficient to ensure that $\operatorname{dim}(V_N)\geq 0$, but is slightly stronger.

For instance, there is no 3-punctured sphere with three $\includegraphics[width=32pt]{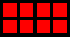}$ punctures. The constraints would overconstrain (imply a negative virtual dimension for) the space of sections of the differential $\phi^{(4)} + 2 \tilde{\phi}$.

\subsection{{Irregular Punctures}}\label{irregular}

In addition to regular punctures, we will, again, need to introduce a class of ``{}irregular''{} punctures, which admit higher-order poles. Ignoring, for the moment, the question of constraints, the class of irregular punctures is the one we introduced in \cite{Chacaltana:2010ks} for the $A_{N-1}$ series\footnote{Our use of the term ``irregular puncture'', in this paper and in \cite{Chacaltana:2010ks}, differs from the conventional one of the Hitchin system literature (e.g., \cite{Gaiotto:2009hg}).}.

\begin{itemize}%
\item Each irregular puncture is associated to a simple subgroup $G\subset Spin(2N)$.

\item From the pole structure $\{p_k\}$, of the irregular puncture, we construct the ``{}conjugate pole structure," $\{p'_k\}$
\begin{itemize}%
\item $p'_k = p_k = k-1$ if $k$ is an exponent of $G$.
\item $p'_k + p_k = 2k-1$ otherwise.
\end{itemize}
\item We demand that the conjugate pole structure be that of a regular puncture, and we denote the irregular puncture, thus constructed, by the Young diagram of the conjugate regular puncture, with one or more ``{}$*$''{}s appended.
\end{itemize}
Incorporating the constraints simply amounts to ``correcting'' which values of $k$ correspond to exponents of $G$.

For example, the $D_4$ puncture, $\includegraphics[width=70pt]{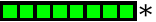}$, has as its conjugate puncture the maximal puncture, $\includegraphics[width=62pt]{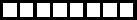}$. Its pole structure, $\{1,3,5;4\}$, allows for a quartic, rather than merely a cubic pole for $\tilde{\phi}$. Thus, the corresponding symmetry group is a $Spin(7)$ subgroup of $Spin(8)$. There are three inequivalent embeddings of $Spin(7)\hookrightarrow Spin(8)$ (depending on which eight-dimensional representation decomposes as the $7+1$). Thus, we also have $\includegraphics[width=70pt]{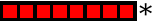}$ and $\includegraphics[width=70pt]{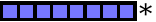}$, which are exchanged by the usual $\mathbb{Z}_2$ outer automorphism. These latter have pole structure $\{1,4,5;4\}$, and impose, respectively, a constraint $c^{(4)}_4\mp 2\tilde{c}_4=0$. This constraint is consequence of using $\phi^{(4)},\, \tilde{\phi}$ as our basis of 4-differentials (rather than the linear combination that appears more naturally at a red/blue puncture).

Similarly, the puncture $\includegraphics[width=77pt]{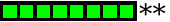}$ corresponds to an $SU(4)$ subgroup of $Spin(8)$, and has poles $\{1,3,6;4\}$. There are again blue and red versions of this puncture corresponding to the other two embeddings of $SU(4)$ related by triality to the green one. The exponent 3 in $SU(4)$ (as opposed to 6) means that we need a constraint $c^{(6)}_6=-(a^{(3)})^2$ that appropriately corrects the dimensions of the Coulomb branch. In a free-field fixture, e.g.,

\begin{displaymath}
 \includegraphics[width=114pt]{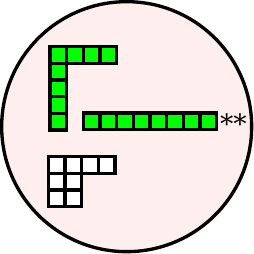}
\end{displaymath}
the constraint $c^{(6)}_6=-(a^{(3)})^2$ from $ \includegraphics[width=77pt]{D4fig15}$ offsets the constraint $c^{(6)}_6=(a^{(3)})^2$ from $ \includegraphics[width=32pt]{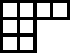}$, so the virtual dimension of the Coulomb branch is indeed equal to its actual dimension (zero).

The red and blue versions of this puncture, $ \includegraphics[width=73pt]{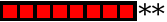}$ and $ \includegraphics[width=73pt]{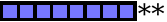}$, have poles $\{1,4,6;4\}$, and have the same constraint as the green one, $c^{(6)}_6=-(a^{(3)})^2$, plus an additional constraint $c^{(4)}_4\mp 2\tilde{c}_4=0$ as usual.

Finally, we can assign a level, $k$, to the $G$ symmetry of the irregular puncture. It is simply defined such that the $G$ gauge group on the cylinder, $p\xleftrightarrow{\quad G\quad}p'$ between $p$ and its conjugate regular puncture $p'$, is conformal.

\subsection{{Central charges}}\label{central}

The conformal-anomaly coefficients, $a$ and $c$, defined via the trace anomaly in a curved background \cite{Osborn:1993cr},

\begin{equation}
\tensor{T}{_{\mu}_{}^{\mu}} = \frac{c}{16\pi^2} {(\text{Weyl})}^2 -\frac{a}{16\pi^2} (\text{Euler}),
\end{equation}
are useful invariants, characterizing 4D conformal field theories. Along with the flavour current-algebra central charges \cite{Argyres:2007cn}, $k_i$, they are among the few readily computable invariants of interacting SCFTs. For the $\mathcal{N}=2$ SCFTs, under discussion, these invariants are constant \cite{Kuzenko:1999pi} over the whole family of SCFTs parametrized by $\mathcal{M}_{g,n}$.

The central charge, $k$, for each simple factor in the flavour symmetry group associated to a regular puncture can be computed directly from the Nahm Young diagram. Denote the length of the $i^{\text{th}}$ row by $r_i$. In the $A_{N-1}$ case, the flavour symmetry group was given by \eqref{AflavourSym} and each $SU(r_i-r_{i+1})$ factor had level

\begin{equation}
k = 2\sum_{j=1}^i r_j
\end{equation}
For the $D_N$ case, the flavour symmetry group is given by \eqref{GflavourDef}, and

\begin{itemize}%
\item For $i$ odd, this gives a ${Spin(r_i-r_{i+1})}_k$ factor in the flavour symmetry group, where

\begin{subequations}
\begin{equation}\label{oddrows}
k = \begin{cases}
   2\left(\sum_{j=1}^i r_j\right)-4& r_i-r_{i+1}\geq 4\\
   4\left(\sum_{j=1}^i r_j\right)-8& r_i-r_{i+1}= 3
 \end{cases}
\end{equation}

\item For $i$ even, this gives an ${Sp\left(\tfrac{r_i-r_{i+1}}{2}\right)}_k$ in the flavour symmetry group, where

\begin{equation}\label{evenrows}
k = \sum_{j=1}^i r_j
\end{equation}
\end{subequations}

\end{itemize}

From Theorem \ref{theorem}, a non-special puncture corresponds to a $2N$-box Nahm Young diagram, whose columns form a D-partition, with at least one (in fact, at least two) odd-length row(s) which appears with odd multiplicity. With a little more work, one can show that at least one of these rows is an even-numbered row. By \eqref{evenrows}, this gives an $Sp(l)_k$ factor, in the flavour symmetry group, with $k$ odd. As mentioned in the introduction, this poses an obstruction to gauging: without additional matter to cancel the anomaly, the $Sp(l)$ gauge theory would suffer from Witten's global anomaly \cite{Witten:1982fp}.

The trace anomaly coefficients, $a$ and $c$, of the SCFT, can be computed (as we did \cite{Chacaltana:2010ks}, for the $A_{N-1}$ series) from two auxiliary quantities: the effective number of hypermultiplets, $n_h$, and the effective number of vector multiplets, $n_v$,

\begin{equation}
\begin{gathered}
a=\tfrac{5n_v+n_h}{24}\\
c=\tfrac{2n_v+n_h}{12}.
\end{gathered}
\label{acnhnv}\end{equation}
In \cite{Chacaltana:2010ks} we gave formul\ae\ to compute $n_h$ and $n_v$ for regular and irregular punctures in the $A_{N-1}$ series. As before, $n_h$ and $n_v$ are the actual number of hypermultiplets and vector multiplets in a \emph{Lagrangian} S-duality frame of the theory, provided such frame exists. As a consequence, the $n_h$ of a free-field fixture (for which $n_v=0$) is equal to the number of free hypermultiplets in this fixture.

To compute $n_v$ for a $D_N$ theory on a curve of genus $g$, one should first calculate the graded dimensions of the Coulomb branch. Then

\begin{equation}
\begin{aligned}
n_v &=\sum_k (2k-1)d_k\\
    &=\sum_{k=1}^{N-1}(4k-1)d_{2k}+\sum_{k=1}^{[\tfrac{N-1}{2}]}(4k+1)d_{2k+1}.
\end{aligned}
\end{equation}
For example, in the $D_4$ theory, the possible non-zero Coulomb branch dimensions are $d_2, d_3, d_4, d_6$, while in the $D_5$ theory, they are $d_2, d_3, d_4, d_5, d_6, d_8$. The odd-degree components of the Coulomb branch of the $D_N$ theory appear only up to degree $2[\tfrac{N-1}{2}]+1$. We will discuss below how to compute the $d_{2k}$ and $d_{2k+1}$, but we will treat the case of $d_{N}$ separately, since it involves the pole orders of the Pfaffian $\tilde{\phi}$.

As we saw before, the even-degree sectors of the Coulomb branch, with dimensions $d_{2k}$ ($2k\neq N$), arise from $2k$-differentials, and so

\begin{equation}
d_{2k} = (1-4k)(1-g) + \sum_{\alpha} (p^{\alpha}_{2k} - s^{\alpha}_{2k} + t^{\alpha}_{2k})
\end{equation}
where $\alpha$ runs over the punctures on the curve, $p^{\alpha}_{2k}$ is the pole order of $\phi_{2k}$ at the $\alpha^{\text{th}}$ puncture, $s^{\alpha}_{2k}$ is the number of constraints of homogeneous degree $2k$ (i.e., polynomial constraints of the form $c^{(2k)}_l=\dots$), and $t^{\alpha}_{2k}$ is the number of $a^{(2k)}$ parameters (i.e., parameters arising from constraints of the form $c^{(4k)}_l=(a^{(2k)})^2$) that the $\alpha^{\text{th}}$ puncture contributes.

On the other hand, since there are no $\phi_{2k+1}$ differentials (except for the Pfaffian, when $N$ is odd), these odd-degree sectors of the Coulomb branch receive contributions \emph{only} from the $a^{(2k+1)}$ parameters (i.e., parameters arising from constraints of the form $c^{(4k+2)}_l=(a^{(2k+1)})^2$). We write

\begin{equation}
d_{2k+1}=\sum_{\alpha} t^{\alpha}_{2k+1},
\end{equation}
Notice that this expression is independent of the genus (in contrast to the contributions, to the $d_{2k}$, from the Riemann-Roch Theorem).

As for $d_N$, if $N$ is even, then $d_N$ gets a contribution from both $\phi_{N}$ and from the Pfaffian $\tilde{\phi}$. The formula for $d_N$ is almost the same as for the $d_{2k}$ case,

\begin{equation}
d_{N} = 2(1-2N)(1-g) + \sum_{\alpha} (p^{\alpha}_{N} - s^{\alpha}_{N})+\tilde{p}^{\alpha}.
\end{equation}
Notice that there is no $t^{\alpha}_{N}$ term, since we do not have a $2N$-differential.

Similarly, if $N$ is odd, only the Pfaffian (the unique odd-degree differential) contributes to $d_N$, and so,

\begin{equation}
d_{N}=(1-2N)(1-g)+\sum_{\alpha}\tilde{p}^{\alpha}.
\end{equation}
Adding up the global, genus-dependent contribution from the $2k$-differentials and the Pfaffian, we obtain

\begin{equation}
n_v=-\tfrac{1}{3}(1-g)N(16N^2-24N+11)+\sum_{\alpha} \delta n_v^{(\alpha)},
\label{nvtot}\end{equation}
where $\alpha$ runs over the punctures on the curve, and the contribution $\delta n_v^{(\alpha)}$ of the $\alpha^{\text{th}}$ puncture to $n_v$ is

\begin{equation}
\delta n_v^{(\alpha)}=\sum_{k=1}^{N-1} (4k-1)(p^{\alpha}_{2k}-s^{\alpha}_{2k}+t^{\alpha}_{2k})
+ \sum_{k=1}^{[\tfrac{N-1}{2}]} (4k+1)t^{\alpha}_{2k+1}
+ (2N-1)\tilde{p}^{\alpha}
\label{nvcontrib}\end{equation}
Let us see a few examples of how to compute $\delta n_v$. First, consider the maximal $D_3$ puncture, which has poles $\{1,3;2\}$, and no constraints. One gets

\begin{equation}
\delta n_v=3(1)+7(3)+5(2)=34.
\end{equation}
Next, consider the $D_4$ puncture, $ \includegraphics[width=32pt]{D4fig4}$. The poles are $\{1,3,4;3\}$ and there is one constraint ($c^{(4)}_3+ 2\tilde{c}_3=0$), so $s_4=1$. We then have

\begin{equation}
\delta n_v=3(1)+7(3-1)+11(4)+7(3)=82.
\end{equation}
Now consider the $D_4$ puncture $ \includegraphics[width=32pt]{D4fig7}$. The poles are $\{1,2,4;2\}$ and there is one constraint ($c^{(6)}_4={\left(a^{(3)}\right)}^2$), so $s_6=1$ and $t_3=1$. Thus,

\begin{equation}
\delta n_v=3(1)+7(2)+11(4-1)+7(2)+5(1)=69.
\end{equation}
Now look at the non-special $D_4$ puncture $ \includegraphics[width=32pt]{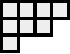}$. Its poles are $\{1,2,4;2\}$, and it has no constraints. This means that

\begin{equation}
\delta n_v=3(1)+7(2)+11(4)+7(2)=75.
\end{equation}
Finally, let us look at the $D_5$ puncture

\begin{equation}
 \includegraphics[width=32pt]{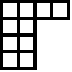}
\end{equation}
which has poles $\{1,2,4,5;3\}$. The two constraints ($c^{(6)}_4=(a^{(3)})^2$ and $c^{(8)}_5=2a^{(3)}\tilde{c}_3$) imply that $t_6=1$, $t_8=1$, and $s_3=1$. Hence,

\begin{equation}
\delta n_v=3(1)+7(2)+11(4-1)+15(5-1)+9(3)+5(1)=142.
\end{equation}
Let us now go on to discuss $n_h$. Just like $n_v$, $n_h$ is a sum of a global piece and contributions from each puncture,

\begin{equation}
n_h=-\tfrac{8}{3}(1-g)N(N-1)(2N-1)+\sum_{\alpha} \delta n_h^{(\alpha)}
\end{equation}
where $\alpha$ runs over the punctures, and

\begin{equation}
\delta n_h^{(\alpha)}=\delta n_v^{(\alpha)}+f^{(\alpha)}
\end{equation}
is the contribution of the $\alpha^{\text{th}}$ puncture to $n_h$. We will see below how to compute $f^{(\alpha)}$ for regular and irregular punctures.

For a \emph{regular} puncture, $f^{(\alpha)}$ can be found\footnote{The contribution $f^{(\text{reg})}=\delta n_h-\delta n_v$ of a regular puncture can be computed from the associated SO-Sp linear quiver tail (as done in \cite{Nanopoulos:2010ga} for the $A_{N-1}$ series), and \eqref{freg} turns out to be, essentially, the dimension \cite{Collingwood1993} of the Nahm (\emph{not} Hitchin) nilpotent orbit. More intrinsically, the individual $\delta n_h$ and $\delta n_v$, rather than their difference, can also be computed from the Nahm nilpotent orbit, as explained in \cite{Chacaltana:2012zy}.} from the row-lengths $r_1\geq r_2\geq \dots$ of the Nahm Young diagram,

\begin{equation}
f^{(\text{reg})}=\frac{1}{4}\sum r_i^2 - \frac{1}{2}\sum r_{\text{odd}},
\label{freg}
\end{equation}
where the first sum is over all rows, and the second is restricted to odd-numbered rows ($r_1, r_3, r_5, r_7, \dots$).

For example, the $D_4$ puncture, $ \includegraphics[width=32pt]{D4fig6}$, has $f=\tfrac{1}{4}[4^2+3^2+1^2]-\tfrac{1}{2}[4+1]=4$. Since we previously computed $n_v=75$ for this puncture, we have $n_h=79$.

The $f^{(\text{irreg})}$ for an irregular puncture, $p$, follows from consistency with degeneration,

\begin{equation}
f^{(\text{irreg})}=-N+\dim G-f^{(\text{reg})},
\end{equation}
where $f^{(\text{reg})}$ is the contribution of the regular puncture, $p'$, conjugate to $p$. $G$ is the flavour symmetry group we ascribe to the irregular puncture, $p$ (equivalently, the gauge group on the cylinder $p\xleftrightarrow{\quad G\quad}p'$).

\subsection{{Regular Punctures (up through $D_6$)}}\label{regular}

We list below the properties of regular punctures for $D_3$, $D_4$, $D_5$, and $D_6$. As explained previously, a puncture in the $D_{2N}$ series is labeled by a Nahm Young diagram, whose column-heights are the parts of a (Nahm) D-partition. On the other hand, the Higgs field boundary condition for the puncture (from which one extracts the pole structure and the constraints), is determined by a Hitchin Young diagram, whose row-lengths are the parts of a (Hitchin) D-partition.

As in the $A_{N-1}$ case, there is a trivial puncture, with Nahm D-partition $[2N-1,1]$ and Hitchin D-partition $[1^{2N}]$ (the zero nilpotent orbit), which corresponds to a non-singular point on the curve $C$, so we exclude it from our discussion.

Also, as already mentioned, for $D_{2N}$, we have red and blue punctures for each very-even D-partition. The constraints for red/blue punctures may differ by a sign. In every case, the top (bottom) sign corresponds to the red (blue) Hitchin D-partition.

Finally, in writing down the global symmetry groups, we find it convenient to use the isomorphisms

\begin{equation}
\begin{gathered}
Spin(2) \simeq U(1)\\
Spin(3) \simeq Sp(1) \simeq SU(2)\\
Spin(4) \simeq {SU(2)}^2\\
Spin(5) \simeq Sp(2)\\
Spin(6) \simeq SU(4)
\end{gathered}
\end{equation}

\subsubsection{{$D_3$}}\label{D3}

Since $D_3\simeq A_3$, the results for $D_3$ were already reported in our previous paper. However, as a warm-up, it will be convenient to repeat them here, recast in the notation we will use for the higher entries in the $D_N$ series.

\bigskip
\noindent
\begin{tabular}{|c|c|c|c|c|c|l|}
\hline 
\mbox{\shortstack{\\Nahm\\YD}}&\mbox{\shortstack{\\Hitchin\\YD}}&\mbox{\shortstack{\\ Pole\\ structure}}&Constraints&\mbox{\shortstack{\\ $A_3$ Nahm\\YD}}&\mbox{\shortstack{\\ Flavour\\ Symmetry}}&$(\delta n_h, \delta n_v)$\\
\hline 
$\begin{matrix} \includegraphics[width=40pt]{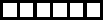}\end{matrix}$&$\begin{matrix} \includegraphics[width=33pt]{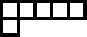}\end{matrix}$&$\{1,3;2\}$&$-$&$\begin{matrix} \includegraphics[width=27pt]{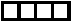}\end{matrix}$&${SU(4)}_8$&$(40,34)$\\
\hline 
$\begin{matrix} \includegraphics[width=27pt]{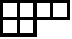}\end{matrix}$&$\begin{matrix} \includegraphics[width=20pt]{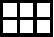}\end{matrix}$&$\{1,2;2\}$&$-$&$\begin{matrix} \includegraphics[width=20pt]{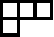}\end{matrix}$&${SU(2)}_6\times U(1)$&$(30,27)$\\
\hline 
$\begin{matrix} \includegraphics[width=27pt]{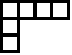}\end{matrix}$&$\begin{matrix} \includegraphics[width=20pt]{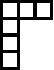}\end{matrix}$&$\{1,2;1\}$&$-$&$\begin{matrix} \includegraphics[width=14pt]{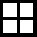}\end{matrix}$&${SU(2)}_8$&$(24,22)$\\
\hline 
$\begin{matrix} \includegraphics[width=14pt]{D3fig4}\end{matrix}$&$\begin{matrix} \includegraphics[width=14pt]{D3fig8}\end{matrix}$&$\{1,2;1\}$&$c^{(4)}_{2}=\tfrac{1}{4}\left(c^{(2)}_{1}\right)^{2}$&$\begin{matrix} \includegraphics[width=14pt]{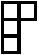}\end{matrix}$&$U(1)$&$(16,15)$\\
\hline 
\end{tabular}
\medskip

Note that, in the $D_3$ description, the quartic differential is allowed to have a double pole at the minimal puncture, instead of only a simple pole (as in the $A_3$ description). However, the coefficient of the double pole is constrained, so that the Coulomb branch has the same graded dimension as before.

\subsubsection{{$D_4$}}\label{D4}

For $D_4$, the outer automorphism group is enhanced from $\mathbb{Z}_2$ to $S_3$. Hence, the pairs of punctures, which were related by exchanging $8_s\leftrightarrow 8_c$, are actually organized into triples, under permutations of $8_s,8_c,8_v$. We indicate this by colouring the Young diagram, corresponding to the other puncture in the triple, green.

The fact that the nilpotent orbits in a triple are related by triality becomes particularly clear if one looks at their weighted Dynkin diagrams \cite{Collingwood1993}. More practical evidence comes from the fact that the punctures in a triple exhibit the same flavour group and $(\delta n_h,\delta n_v)$.

In this table, and in the $D_5$, $D_6$ tables below, we have shaded each non-special Nahm Young diagram and the (special) Hitchin Young diagram which is its image under the Spaltenstein map.
\bigskip

\noindent
\begin{tabular}{|c|c|c|c|c|l|}
\hline
\mbox{\shortstack{\\Nahm\\YD}}&\mbox{\shortstack{\\Hitchin\\YD}}&\mbox{\shortstack{\\ Pole\\ structure}}&Constraints&\mbox{\shortstack{\\ Flavour\\ Symmetry}}&$(\delta n_h, \delta n_v)$\\
\hline 
$\begin{matrix} \includegraphics[width=62pt]{D4fig1}\end{matrix}$&$\begin{matrix} \includegraphics[width=54pt]{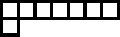}\end{matrix}$&$\{1,3,5;3\}$&$-$&${Spin(8)}_{12}$&$(112,100)$\\
\hline
$\begin{matrix} \includegraphics[width=47pt]{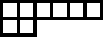}\end{matrix}$&$\begin{matrix} \includegraphics[width=39pt]{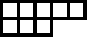}\end{matrix}$&$\{1,3,4;3\}$&$-$&${SU(2)}_{8}^3$&$(96,89)$\\
\hline
$\begin{matrix} \includegraphics[width=47pt]{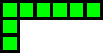}\end{matrix}$&$\begin{matrix} \includegraphics[width=39pt]{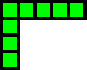}\end{matrix}$&$\{1,3,4;2\}$&$-$&${Sp(2)}_8$&$(88,82)$\\
\hline
$\begin{array}{l} \includegraphics[width=32pt]{D4fig4}\, ,\\  \includegraphics[width=32pt]{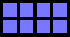}\end{array}$&$\begin{array}{l} \includegraphics[width=32pt]{D4fig4}\, ,\\  \includegraphics[width=32pt]{D4fig5}\end{array}$&$\{1,3,4;3\}$&$c^{(4)}_{3}\pm 2\tilde{c}_{3}=0$&${Sp(2)}_8$&$(88,82)$\\
\hline
$\begin{matrix} \includegraphics[width=32pt]{D4fig7}\end{matrix}$&$\begin{matrix} \includegraphics[width=24pt]{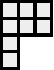}\end{matrix}$&$\{1,2,4;2\}$&$c^{(6)}_{4}={\left(a^{(3)}\right)}^2$&${U(1)}^2$&$(72,69)$\\
\hline
$\begin{matrix} \includegraphics[width=32pt]{D4fig6}\end{matrix}$&$\begin{matrix} \includegraphics[width=24pt]{D4fig245}\end{matrix}$&$\{1,2,4;2\}$&$-$&${SU(2)}_7$&$(79,75)$\\
\hline
$\begin{matrix} \includegraphics[width=32pt]{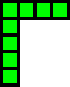}\end{matrix}$&$\begin{matrix} \includegraphics[width=24pt]{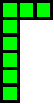}\end{matrix}$&$\{1,2,2;1\}$&$-$&${SU(2)}_8$&$(48,46)$\\
\hline
$\begin{array}{l} \includegraphics[width=17pt]{D4fig9}\, ,\,  \includegraphics[width=17pt]{D4fig10}\end{array}$&$\begin{array}{l} \includegraphics[width=17pt]{D4fig9}\, ,\,  \includegraphics[width=17pt]{D4fig10}\end{array}$&$\{1,2,3;2\}$&$\begin{gathered}c^{(4)}_{2}\pm 2\tilde{c}_{2}=\tfrac{1}{4}\left(c^{(2)}_{1}\right)^2\\ c^{(6)}_3=\mp \tilde{c}_2 c^{(2)}_{1} \end{gathered}$&${SU(2)}_8$&$(48,46)$\\
\hline
$\begin{matrix} \includegraphics[width=17pt]{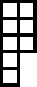}\end{matrix}$&$\begin{matrix} \includegraphics[width=17pt]{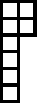}\end{matrix}$&$\{1,2,2;1\}$&$c^{(4)}_2=\tfrac{1}{4}\left(c^{(2)}_1\right)^2$&none&$(40,39)$\\
\hline
\end{tabular}

\subsubsection{{$D_5$}}\label{D5}

{\parindent=-0.25in
\noindent
\begin{tabular}{|c|c|c|c|c|l|}
\hline
\mbox{\shortstack{\\Nahm\\YD}}&\mbox{\shortstack{\\Hitchin\\YD}}&\mbox{\shortstack{\\ Pole\\ structure}}&Constraints&\mbox{\shortstack{\\ Flavour\\ Symmetry}}&$(\delta n_h, \delta n_v)$\\
\hline 
$\begin{matrix} \includegraphics[width=58pt]{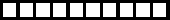}\end{matrix}$&$\begin{matrix} \includegraphics[width=52pt]{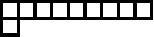}\end{matrix}$&$\{1,3,5,7;4\}$&$-$&${Spin(10)}_{16}$&$(240,220)$\\
\hline
$\begin{matrix} \includegraphics[width=47pt]{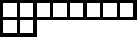}\end{matrix}$&$\begin{matrix} \includegraphics[width=41pt]{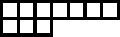}\end{matrix}$&$\{1,3,5,6;4\}$&$-$&${SU(4)}_{12}\times {SU(2)}_{10}$&$(218,205)$\\
\hline
\end{tabular}

\noindent
\begin{tabular}{|c|c|c|c|c|l|}
\hline
\mbox{\shortstack{\\Nahm\\YD}}&\mbox{\shortstack{\\Hitchin\\YD}}&\mbox{\shortstack{\\ Pole\\ structure}}&Constraints&\mbox{\shortstack{\\ Flavour\\ Symmetry}}&$(\delta n_h, \delta n_v)$\\
\hline
$\begin{matrix} \includegraphics[width=47pt]{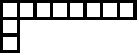}\end{matrix}$&$\begin{matrix} \includegraphics[width=41pt]{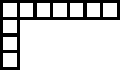}\end{matrix}$&$\{1,3,5,6;3\}$&$-$&${Spin(7)}_{12}$&$(208,196)$\\
\hline
$\begin{matrix} \includegraphics[width=35pt]{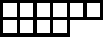}\end{matrix}$&$\begin{matrix} \includegraphics[width=29pt]{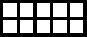}\end{matrix}$&$\{1,3,4,6;4\}$&$-$&${Sp(2)}_{10}\times U(1)$&$(204,194)$\\
\hline
$\begin{matrix} \includegraphics[width=35pt]{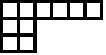}\end{matrix}$&$\begin{matrix} \includegraphics[width=29pt]{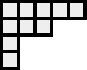}\end{matrix}$&$\{1,3,4,6;3\}$&$c^{(8)}_6 = {\left(a^{(4)}\right)}^2$&${SU(2)}_8^2\times U(1)$&$(184,177)$\\
\hline
$\begin{matrix} \includegraphics[width=35pt]{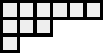}\end{matrix}$&$\begin{matrix} \includegraphics[width=29pt]{D4fig252}\end{matrix}$&$\{1,3,4,6;3\}$&$-$&${SU(2)}_{16}\times {SU(2)}_9$&$(193,185)$\\
\hline
$\begin{matrix} \includegraphics[width=24pt]{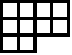}\end{matrix}$&$\begin{matrix} \includegraphics[width=24pt]{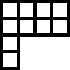}\end{matrix}$&$\{1,3,4,6;3\}$&$c^{(8)}_6=\tfrac{1}{4}{\left(c^{(4)}_3\right)}^2$&${SU(2)}_8\times U(1)$&$(176,170)$\\
\hline
$\begin{array}{l} \includegraphics[width=24pt]{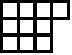}\end{array}$&$\begin{matrix} \includegraphics[width=18pt]{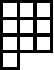}\end{matrix}$&$\{1,2,4,5;3\}$&$-$&${SU(2)}_{32}$&$(168,163)$\\
\hline
$\begin{matrix} \includegraphics[width=35pt]{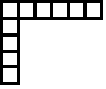}\end{matrix}$&$\begin{matrix} \includegraphics[width=29pt]{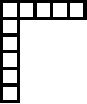}\end{matrix}$&$\{1,3,4,4;2\}$&$-$&${Sp(2)}_8$&$(152,146)$\\
\hline
$\begin{matrix} \includegraphics[width=24pt]{D5fig9}\end{matrix}$&$\begin{matrix} \includegraphics[width=18pt]{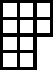}\end{matrix}$&$\{1,2,4,5;3\}$&$\begin{gathered}c^{(6)}_4 = {\bigl(a^{(3)}\bigr)}^2\\ c^{(8)}_5=2 a^{(3)} \tilde{c}_3\end{gathered}$&${SU(2)}_{10}\times U(1)$&$(146,142)$\\
\hline
$\begin{matrix} \includegraphics[width=24pt]{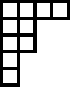}\end{matrix}$&$\begin{matrix} \includegraphics[width=18pt]{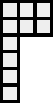}\end{matrix}$&$\{1,2,4,4;2\}$&$c^{(6)}_4 = {\bigl(a^{(3)}\bigr)}^2$&$U(1)$&$(136,133)$\\
\hline
$\begin{matrix} \includegraphics[width=24pt]{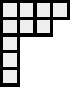}\end{matrix}$&$\begin{matrix} \includegraphics[width=18pt]{D4fig258}\end{matrix}$&$\{1,2,4,4;2\}$&$-$&${SU(2)}_7$&$(143,139)$\\
\hline
$\begin{matrix} \includegraphics[width=13pt]{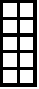}\end{matrix}$&$\begin{matrix} \includegraphics[width=13pt]{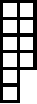}\end{matrix}$&$\{1,2,3,4;2\}$&
$\begin{gathered}
c'^{(4)}_2 \equiv c^{(4)}_2 - \tfrac{1}{4} {\bigl(c^{(2)}_1\bigr)}^2\\
c^{(6)}_3 = \tfrac{1}{2} c^{(2)}_1 c'^{(4)}_2\\
c^{(8)}_4 = \tfrac{1}{4} {\bigl( c'^{(4)}_2\bigr)}^2\end{gathered}$
&$U(1)$&$(104,102)$\\
\hline
$\begin{array}{l} \includegraphics[width=24pt]{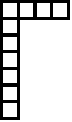}\end{array}$&$\begin{matrix} \includegraphics[width=18pt]{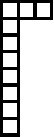}\end{matrix}$&$\{1,2,2,2;1\}$&$-$&${SU(2)}_8$&$(80,78)$\\
\hline
$\begin{matrix} \includegraphics[width=13pt]{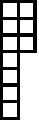}\end{matrix}$&$\begin{matrix} \includegraphics[width=13pt]{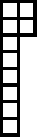}\end{matrix}$&$\{1,2,2,2;1\}$&$c^{(4)}_{2}=\tfrac{1}{4}{\bigl(c^{(2)}_{1}\bigr)}^2$&none&$(72,71)$\\
\hline
\end{tabular}
}

\subsubsection{{$D_6$}}\label{D6}

{\parindent=-0.25in
\noindent
\begin{tabular}{|c|c|c|c|c|l|}
\hline
\mbox{\shortstack{\\Nahm\\YD}}&\mbox{\shortstack{\\Hitchin\\YD}}&Pole structure&Constraints&\mbox{\shortstack{\\ Flavour\\ Symmetry}}&$(\delta n_h, \delta n_v)$\\
\hline 
$\begin{matrix} \includegraphics[width=59pt]{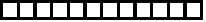}\end{matrix}$&$\begin{matrix} \includegraphics[width=54pt]{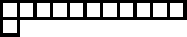}\end{matrix}$&$\{1,3,5,7,9;5\}$&$-$&${Spin(12)}_{20}$&$(440,410)$\\
\hline
$\begin{matrix} \includegraphics[width=49pt]{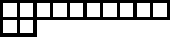}\end{matrix}$&$\begin{matrix} \includegraphics[width=44pt]{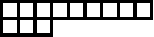}\end{matrix}$&$\{1,3,5,7,8;5\}$&$-$&\small${Spin(8)}_{16}\times {SU(2)}_{12}$&$(412,391)$\\
\hline
$\begin{matrix} \includegraphics[width=49pt]{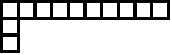}\end{matrix}$&$\begin{matrix} \includegraphics[width=44pt]{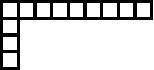}\end{matrix}$&$\{1,3,5,7,8;4\}$&$-$&${Spin(9)}_{16}$&$(400,380)$\\
\hline
\end{tabular}

\noindent
\begin{tabular}{|c|c|c|c|c|l|}
\hline
\mbox{\shortstack{\\Nahm\\YD}}&\mbox{\shortstack{\\Hitchin\\YD}}&Pole structure&Constraints&\mbox{\shortstack{\\ Flavour\\ Symmetry}}&$(\delta n_h, \delta n_v)$\\
\hline
$\begin{matrix} \includegraphics[width=40pt]{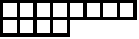}\end{matrix}$&$\begin{matrix} \includegraphics[width=35pt]{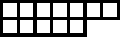}\end{matrix}$&$\{1,3,5,6,8;5\}$&$-$&\small${Sp(2)}_{12}\times {SU(2)}_{12}^2$&$(392,376)$\\
\hline
$\begin{array}{l} \includegraphics[width=30pt]{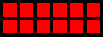}\, ,\\  \includegraphics[width=30pt]{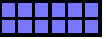}\end{array}$&$\begin{array}{l} \includegraphics[width=30pt]{D6fig5}\, ,\\  \includegraphics[width=30pt]{D6fig6}\end{array}$&$\{1,3,5,6,8;5\}$&$c^{(6)}_5\pm 2\tilde{c}_5=0$&${Sp(3)}_{12}$&$(380,365)$\\
\hline
$\begin{matrix} \includegraphics[width=40pt]{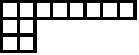}\end{matrix}$&$\begin{matrix} \includegraphics[width=35pt]{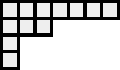}\end{matrix}$&$\{1,3,5,6,8;4\}$&$c^{(10)}_8={(a^{(5)})}^2$&${SU(4)}_{12}\times U(1)$&$(368,355)$\\
\hline
$\begin{matrix} \includegraphics[width=40pt]{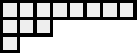}\end{matrix}$&$\begin{matrix} \includegraphics[width=35pt]{D4fig268}\end{matrix}$&$\{1,3,5,6,8;4\}$&$-$&${Sp(2)}_{12}\times {SU(2)}_{11}$&$(379,365)$\\
\hline
$\begin{matrix} \includegraphics[width=30pt]{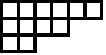}\end{matrix}$&$\begin{matrix} \includegraphics[width=25pt]{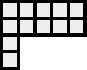}\end{matrix}$&$\{1,3,4,6,8;4\}$&$c^{(10)}_8={(a^{(5)})}^2$&${SU(2)}_{10}\times U(1)^2$&$(354,344)$\\
\hline
$\begin{matrix} \includegraphics[width=30pt]{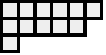}\end{matrix}$&$\begin{matrix} \includegraphics[width=25pt]{D4fig270}\end{matrix}$&$\{1,3,4,6,8;4\}$&$-$&${Sp(2)}_{11}$&$(366,354)$\\
\hline
$\begin{matrix} \includegraphics[width=30pt]{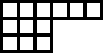}\end{matrix}$&$\begin{matrix} \includegraphics[width=25pt]{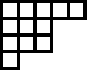}\end{matrix}$&$\{1,3,4,6,7;4\}$&$-$&${SU(2)}_{40}\times {SU(2)}_{16}$&$(344,335)$\\
\hline
$\begin{matrix} \includegraphics[width=20pt]{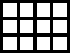}\end{matrix}$&$\begin{matrix} \includegraphics[width=20pt]{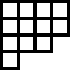}\end{matrix}$&$\{1,3,4,6,7;4\}$&$c^{(8)}_6 = \tfrac{1}{4}{(c^{(4)}_3)}^2$&${SU(2)}_{20}^2$&$(328,320)$\\
\hline
$\begin{matrix} \includegraphics[width=40pt]{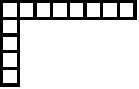}\end{matrix}$&$\begin{matrix} \includegraphics[width=35pt]{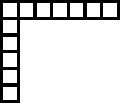}\end{matrix}$&$\{1,3,5,6,6;3\}$&$-$&${Spin(7)}_{12}$&$(328,316)$\\
\hline
$\begin{matrix} \includegraphics[width=30pt]{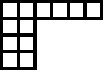}\end{matrix}$&$\begin{matrix} \includegraphics[width=25pt]{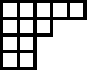}\end{matrix}$&$\{1,3,4,6,7;4\}$&$\begin{gathered}c^{(8)}_6= {(a^{(4)})}^2\\ c^{(10)}_7=a^{(4)}\tilde{c}_4\end{gathered}$&${SU(2)}_{12}\times {SU(2)}_8^2$&$(316,308)$\\
\hline
$\begin{array}{l} \includegraphics[width=20pt]{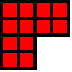}\, ,\,  \includegraphics[width=20pt]{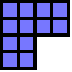}\end{array}$&$\begin{array}{l} \includegraphics[width=20pt]{D6fig16}\, ,\,  \includegraphics[width=20pt]{D6fig17}\end{array}$&$\{1,3,4,6,7;4\}$&$\begin{gathered}c^{(8)}_6=\tfrac{1}{4}{(c^{(4)}_3)}^2\\ c^{(10)}_7 =\pm\tilde{c}_4 c^{(4)}_3 \end{gathered}$&${SU(2)}_{12}\times{SU(2)}_8$&$(308,301)$\\
\hline
$\begin{matrix} \includegraphics[width=30pt]{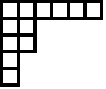}\end{matrix}$&$\begin{matrix} \includegraphics[width=25pt]{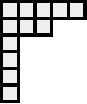}\end{matrix}$&$\{1,3,4,6,6;3\}$&$c^{(8)}_6={\left(a^{(4)}\right)}^2$&${SU(2)}_8^2$&$(304,297)$\\
\hline
$\begin{matrix} \includegraphics[width=30pt]{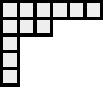}\end{matrix}$&$\begin{matrix} \includegraphics[width=25pt]{D4fig275}\end{matrix}$&$\{1,3,4,6,6;3\}$&$-$&${SU(2)}_{16}\times {SU(2)}_9$&$(313,305)$\\
\hline
$\begin{matrix} \includegraphics[width=20pt]{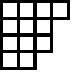}\end{matrix}$&$\begin{matrix} \includegraphics[width=15pt]{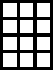}\end{matrix}$&$\{1,2,4,5,6;4\}$&$-$&${SU(2)}_{12}$&$(300,294)$\\
\hline
$\begin{matrix} \includegraphics[width=20pt]{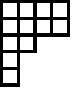}\end{matrix}$&$\begin{matrix} \includegraphics[width=20pt]{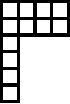}\end{matrix}$&$\{1,3,4,6,6;3\}$&$c^{(8)}_6=\tfrac{1}{4}{(c^{(4)}_3)}^2$&${SU(2)}_8$&$(296,290)$\\
\hline
$\begin{matrix} \includegraphics[width=20pt]{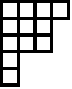}\end{matrix}$&$\begin{matrix} \includegraphics[width=15pt]{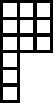}\end{matrix}$&$\{1,2,4,5,6;3\}$&$-$&$U(1)$&$(288,283)$\\
\hline
$\begin{matrix} \includegraphics[width=20pt]{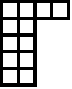}\end{matrix}$&$\begin{matrix} \includegraphics[width=15pt]{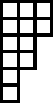}\end{matrix}$&$\{1,2,4,5,6;3\}$&$\begin{gathered}c^{(6)}_4={\left(a^{(3)}\right)}^2\\ c^{(10)}_6={\left(a^{(5)}\right)}^2\\ c^{(8)}_4 = 2 a^{(3)} a^{(5)} \end{gathered}$&$U(1)^2$&$(256,252)$\\
\hline
$\begin{matrix} \includegraphics[width=30pt]{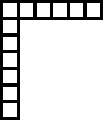}\end{matrix}$&$\begin{matrix} \includegraphics[width=25pt]{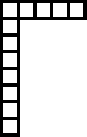}\end{matrix}$&$\{1,3,4,4,4;2\}$&$-$&${Sp(2)}_8$&$(232,226)$\\
\hline
$\begin{matrix} \includegraphics[width=20pt]{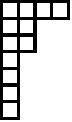}\end{matrix}$&$\begin{matrix} \includegraphics[width=15pt]{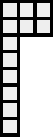}\end{matrix}$&$\{1,2,4,4,4;2\}$&$c^{(6)}_4={\left(a^{(3)}\right)}^2$&$U(1)$&$(216,213)$\\
\hline
\end{tabular}

\noindent
\begin{tabular}{|c|c|c|c|c|l|}
\hline
\mbox{\shortstack{\\Nahm\\YD}}&\mbox{\shortstack{\\Hitchin\\YD}}&Pole structure&Constraints&\mbox{\shortstack{\\ Flavour\\ Symmetry}}&$(\delta n_h, \delta n_v)$\\
\hline
$\begin{matrix} \includegraphics[width=20pt]{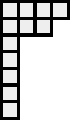}\end{matrix}$&$\begin{matrix} \includegraphics[width=15pt]{D4fig281}\end{matrix}$&$\{1,2,4,4,4;2\}$&$-$&${SU(2)}_7$&$(223,219)$\\
\hline
$\begin{array}{l} \includegraphics[width=11pt]{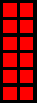}\, ,\,  \includegraphics[width=11pt]{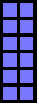}\end{array}$&$\begin{array}{l} \includegraphics[width=11pt]{D6fig25}\, ,\,  \includegraphics[width=11pt]{D6fig26}\end{array}$&$\{1,2,3,4,5;3\}$&
$\begin{gathered}
c'^{(4)}_2\equiv c^{(4)}_2-\tfrac{1}{4}{\bigl(c^{(2)}_1\bigr)}^2\\
c^{(6)}_3\mp 2\tilde{c}_3 =\tfrac{1}{2} c^{(2)}_1 c'^{(4)}_2\\ c^{(8)}_4 = \tfrac{1}{4} {\Bigl(c'^{(4)}_2\Bigr)}^2\pm \tilde{c}_3 c^{(2)}_1\\ c^{(10)}_5 = \pm \tilde{c}_3 c'^{(4)}_2 \end{gathered}$&${SU(2)}_{12}$&$(196,193)$\\
\hline
$\begin{matrix} \includegraphics[width=11pt]{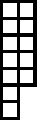}\end{matrix}$&$\begin{matrix} \includegraphics[width=11pt]{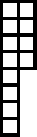}\end{matrix}$&$\{1,2,3,4,4;2\}$&
$\begin{gathered}
c'^{(4)}_2\equiv c^{(4)}_2-\tfrac{1}{4}{\bigl(c^{(2)}_1\bigr)}^2\\ 
c^{(6)}_3 = \tfrac{1}{2} c^{(2)}_1 c'^{(4)}_2\\
c^{(8)}_4 = \tfrac{1}{4}{\Bigl(c'^{(4)}_2\Bigr)}^2\end{gathered}$
&none&$(184,182)$\\
\hline
$\begin{matrix} \includegraphics[width=20pt]{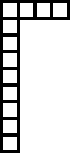}\end{matrix}$&$\begin{matrix} \includegraphics[width=15pt]{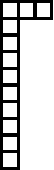}\end{matrix}$&$\{1,2,2,2,2;1\}$&$-$&${SU(2)}_8$&$(120,118)$\\
\hline
$\begin{matrix} \includegraphics[width=11pt]{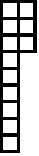}\end{matrix}$&$\begin{matrix} \includegraphics[width=11pt]{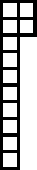}\end{matrix}$&$\{1,2,2,2,2;1\}$&$c^{(4)}_2 =\tfrac{1}{4}{\left(c^{(2)}_1\right)}^2$&none&$(112,111)$\\
\hline
\end{tabular}
}

\section{{The $D_4$ theory}}\label{D4Theory}

In this section, we will develop the complete ``{}tinkertoy''{} catalogue for the $D_4$ theory. The regular punctures are listed in \S\ref{D4}. Every irregular puncture arises from the collision of a pair of regular punctures. Since there exist cylinders that connect two irregular punctures, it is possible to find (as we did in \cite{Chacaltana:2010ks} for the $A_{N-1}$ series) the full list of irregular punctures, cylinders and fixtures by considering the degenerations of all 4-punctured spheres that are ``good'' (i.e., that have non-negative Coulomb branch dimensions \cite{Gaiotto:2008ak,Gaiotto:2011xs}). In the end, our fixtures that include an irregular puncture are ``ugly'' (and typically include a number of free hypers), while those which do not are ``good''. It is possible that at least some ``bad'' punctured Riemann surfaces possess a sensible 4D $\mathcal{N}=2$ low-energy interpretation, as stressed in \cite{Gaiotto:2011xs}, but we do not attempt to cover them in this paper.

\subsection{{Irregular punctures and cylinders}}\label{D4irregular}

For irregular punctures, we show the Nahm Young diagram of their conjugate regular puncture. The number of stars accompanying the Nahm Young diagram simply serves to enumerate the distinct irregular punctures with the same conjugate regular puncture.

\noindent
\begin{tabular}{|l|c|c|c|l|}
\hline
Nahm YD&Pole structure&Constraints&Flavour Symmetry&$(\delta n_h, \delta n_v)$\\
\hline 
$\begin{array}{l} \includegraphics[width=70pt]{D4fig12}\end{array}$&$\{1,3,5;4\}$&$-$&${Spin(7)}_8$&$(112,107)$\\
\hline
$\begin{array}{l} \includegraphics[width=70pt]{D4fig13}\, ,\\  \includegraphics[width=70pt]{D4fig14}\end{array}$&$\{1,4,5;4\}$&$c^{(4)}_4\mp 2\tilde{c}_4=0$&${Spin(7)}_8$&$(112,107)$\\
\hline 
$\begin{array}{l} \includegraphics[width=77pt]{D4fig15}\end{array}$&$\{1,3,6;4\}$&$c^{(6)}_6= -{(a^{(3)})}^2$&${SU(4)}_4$&$(112,113)$\\
\hline 
$\begin{array}{l} \includegraphics[width=73pt]{D4fig16}\, ,\\  \includegraphics[width=75pt]{D4fig17}\end{array}$&$\{1,4,6;4\}$&$\begin{gathered}c^{(4)}_4\mp 2\tilde{c}_4=0\\ c^{(6)}_6= -{(a^{(3)})}^2\end{gathered}$&${SU(4)}_4$&$(112,113)$\\
\hline 
$\begin{array}{l} \includegraphics[width=80pt]{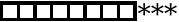}\end{array}$&$\{1,4,5;4\}$&$-$&${(G_2)}_4$&$(112,114)$\\
\hline 
$\begin{array}{l} \includegraphics[width=89pt]{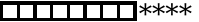}\end{array}$&$\{1,4,6;4\}$&$c^{(6)}_6= -{(a^{(3)})}^2$&${SU(3)}_0$&$(112,120)$\\
\hline 
$\begin{array}{l} \includegraphics[width=55pt]{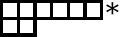}\end{array}$&$\{1,4,7;4\}$&$-$&${SU(2)}_0$&$(128,136)$\\
\hline 
$\begin{array}{l} \includegraphics[width=56pt]{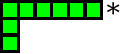}\end{array}$&$\{1,3,7;5\}$&$-$&${Sp(2)}_4$&$(136,136)$\\
\hline 
$\begin{array}{l} \includegraphics[width=40pt]{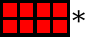}\, ,\\  \includegraphics[width=40pt]{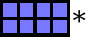}\end{array}$&$\{1,5,7;5\}$&$\begin{gathered}c^{(4)}_5\mp\tilde{c}_5=0\\ c^{(4)}_4\mp\tilde{c}_4=0\end{gathered}$&${Sp(2)}_4$&$(136,136)$\\
\hline 
$\begin{array}{l} \includegraphics[width=59pt]{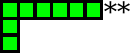}\end{array}$&$\{1,4,7;5\}$&$-$&${SU(2)}_0$&$(136,143)$\\
\hline 
$\begin{array}{l} \includegraphics[width=49pt]{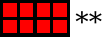}\, ,\\  \includegraphics[width=49pt]{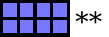}\end{array}$&$\{1,5,7;5\}$&$c^{(4)}_5\mp\tilde{c}_5=0$&${SU(2)}_0$&$(136,143)$\\
\hline
$\begin{array}{l} \includegraphics[width=39pt]{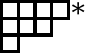}\end{array}$&$\{1,5,7;5\}$&$-$&${SU(2)}_1$&$(145,150)$\\
\hline 
\end{tabular}
\vfill

The cylinders in the $D_4$ theory are

\begin{displaymath}
\begin{gathered}
\begin{matrix} \includegraphics[width=62pt]{D4fig1}\end{matrix}
\xleftrightarrow{\qquad Spin(8)\qquad}
\begin{matrix} \includegraphics[width=62pt]{D4fig1}\end{matrix}\\
\begin{matrix} \includegraphics[width=62pt]{D4fig1}\end{matrix}
\xleftrightarrow{\qquad Spin(7)\qquad}
\begin{matrix} \includegraphics[width=70pt]{D4fig12}\end{matrix}\\
\begin{matrix} \includegraphics[width=62pt]{D4fig1}\end{matrix}
\xleftrightarrow{\qquad Spin(7)\qquad}
\begin{matrix}
 \includegraphics[width=70pt]{D4fig13}\end{matrix}\\
\begin{matrix} \includegraphics[width=62pt]{D4fig1}\end{matrix}
\xleftrightarrow{\qquad Spin(7)\qquad}
\begin{matrix}
 \includegraphics[width=70pt]{D4fig14}\end{matrix}\\
\begin{matrix} \includegraphics[width=62pt]{D4fig1}\end{matrix}
\xleftrightarrow{\qquad SU(4)\qquad}
\begin{matrix} \includegraphics[width=77pt]{D4fig15}\end{matrix}\\
\begin{matrix} \includegraphics[width=62pt]{D4fig1}\end{matrix}
\xleftrightarrow{\qquad SU(4)\qquad}
\begin{matrix} \includegraphics[width=75pt]{D4fig16}\end{matrix}\\
\begin{matrix} \includegraphics[width=62pt]{D4fig1}\end{matrix}
\xleftrightarrow{\qquad SU(4)\qquad}
\begin{matrix} \includegraphics[width=75pt]{D4fig17}\end{matrix}\\
\begin{matrix} \includegraphics[width=62pt]{D4fig1}\end{matrix}
\xleftrightarrow{\qquad G_2\qquad}
\begin{matrix} \includegraphics[width=80pt]{D4fig18}\end{matrix}\\
\begin{matrix} \includegraphics[width=70pt]{D4fig12}\end{matrix}
\xleftrightarrow{\qquad G_2\qquad}
\begin{matrix} \includegraphics[width=70pt]{D4fig13}\end{matrix}\\
\begin{matrix} \includegraphics[width=70pt]{D4fig14}\end{matrix}\xleftrightarrow{\qquad G_2\qquad}\begin{matrix} \includegraphics[width=70pt]{D4fig12}\end{matrix}\\
\begin{matrix} \includegraphics[width=70pt]{D4fig13}\end{matrix}\xleftrightarrow{\qquad G_2\qquad}\begin{matrix} \includegraphics[width=77pt]{D4fig14}\end{matrix}\\
\end{gathered}
\end{displaymath}

\begin{displaymath}
\begin{gathered}
\begin{matrix} \includegraphics[width=70pt]{D4fig12}\end{matrix}
\xleftrightarrow{\qquad SU(3)\qquad}
\begin{matrix} \includegraphics[width=75pt]{D4fig16}\end{matrix}\\
\begin{matrix} \includegraphics[width=70pt]{D4fig12}\end{matrix}
\xleftrightarrow{\qquad SU(3)\qquad}
\begin{matrix} \includegraphics[width=75pt]{D4fig17}\end{matrix}\\
\begin{matrix}
 \includegraphics[width=70pt]{D4fig13}\end{matrix}
\xleftrightarrow{\qquad SU(3)\qquad}
\begin{matrix} \includegraphics[width=77pt]{D4fig15}\end{matrix}\\
\begin{matrix}
 \includegraphics[width=70pt]{D4fig13}\end{matrix}
\xleftrightarrow{\qquad SU(3)\qquad}
\begin{matrix} \includegraphics[width=75pt]{D4fig17}\end{matrix}\\
\begin{matrix}
 \includegraphics[width=70pt]{D4fig14}\end{matrix}
\xleftrightarrow{\qquad SU(3)\qquad}
\begin{matrix} \includegraphics[width=77pt]{D4fig15}\end{matrix}\\
\begin{matrix}
 \includegraphics[width=70pt]{D4fig14}\end{matrix}\xleftrightarrow{\qquad SU(3)\qquad}
\begin{matrix} \includegraphics[width=75pt]{D4fig16}\end{matrix}\\
\begin{matrix} \includegraphics[width=62pt]{D4fig1}\end{matrix}
\xleftrightarrow{\qquad SU(3)\qquad}
\begin{matrix} \includegraphics[width=89pt]{D4fig27}\end{matrix}\\
\begin{matrix} \includegraphics[width=47pt]{D4fig3}\end{matrix}
\xleftrightarrow{\qquad Sp(2)\qquad}
\begin{matrix} \includegraphics[width=56pt]{D4fig20}\end{matrix}\\
\begin{matrix} \includegraphics[width=32pt]{D4fig4}\end{matrix}
\xleftrightarrow{\qquad Sp(2)\qquad}
\begin{matrix} \includegraphics[width=40pt]{D4fig21}\end{matrix}\\
\begin{matrix} \includegraphics[width=32pt]{D4fig5}\end{matrix}
\xleftrightarrow{\qquad Sp(2)\qquad}
\begin{matrix} \includegraphics[width=40pt]{D4fig22}\end{matrix}\\
\begin{matrix} \includegraphics[width=47pt]{D4fig3}\end{matrix}
\xleftrightarrow{\qquad SU(2)\qquad}
\begin{matrix} \includegraphics[width=59pt]{D4fig23}\end{matrix}\\
\begin{matrix} \includegraphics[width=32pt]{D4fig4}\end{matrix}
\xleftrightarrow{\qquad SU(2)\qquad}
\begin{matrix} \includegraphics[width=49pt]{D4fig24}\end{matrix}\\
\begin{matrix} \includegraphics[width=32pt]{D4fig5}\end{matrix}
\xleftrightarrow{\qquad SU(2)\qquad}
\begin{matrix} \includegraphics[width=49pt]{D4fig25}\end{matrix}\\
\begin{matrix} \includegraphics[width=47pt]{D4fig2}\end{matrix}
\xleftrightarrow{\qquad SU(2)\qquad}
\begin{matrix} \includegraphics[width=55pt]{D4fig19}\end{matrix}\\
\begin{matrix} \includegraphics[width=32pt]{D4fig6}\end{matrix}
\xleftrightarrow{\qquad SU(2)\qquad}
\begin{matrix} \includegraphics[width=39pt]{D4fig26}\end{matrix}
\end{gathered}
\end{displaymath}
\medskip

Note that some of the irregular punctures have level $k=0$. Appropriately, these will appear, below, on ``{}empty''{} fixtures, with zero hypermultiplets. Also, note that each of the cylinders, $p\xleftrightarrow{\quad G \quad}p'$, satisfies

\begin{equation}
\begin{aligned}
\delta n_h + \delta {n_h}' -  8N(N-1)(2N-1)/3 &=0\\
\delta n_v + \delta {n_v}' - N(16N^2-24N+11)/3 & = dim(G)\\
k+k' = k_{\text{critical}}
\end{aligned}
\end{equation}
where $k_{\text{critical}}=2\ell_{\text{adj}}$ is the value of $k$ which gives vanishing $\beta$-function for $G$. While this was true (by construction) when $p'$ is the conjugate regular puncture to $p$, it is not automatically-satisfied for cylinders between two irregular punctures. In essence, these conditions determine which cylinders between pairs of irregular punctures are allowed.

\subsection{{Fixtures}}\label{D4fixtures}

Here, we list all of the 3-punctured spheres. There are a lot of them, but fortunately, the profusion is partially tamed by the fact that they are organized into multiplets under the outer automorphism group.

\subsubsection{{Free-field fixtures}}\label{D4free}

Free-field fixtures are either empty, or contain only free matter hypermultiplets, in some representation of the global symmetry group for the fixture. Below, we show the matter representations only for the non-Abelian part of the global symmetry group.
\bigskip

\noindent
\begin{tabular}{|c|c|c|}
\hline 
Fixture&\mbox{\shortstack{\\ Number\\ of Hypers}}&Representation\\
\hline 
$\begin{matrix} \includegraphics[width=64pt]{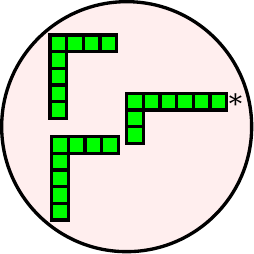}\end{matrix},\, \begin{matrix} \includegraphics[width=64pt]{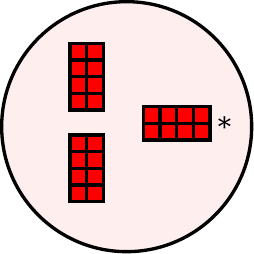}\end{matrix},\, \begin{matrix} \includegraphics[width=64pt]{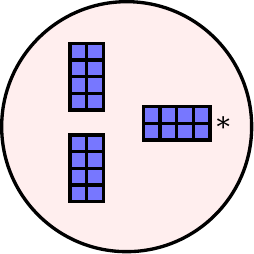}\end{matrix}$&8&$\tfrac{1}{2}(2,2,4)$\\
\hline
$\begin{matrix} \includegraphics[width=64pt]{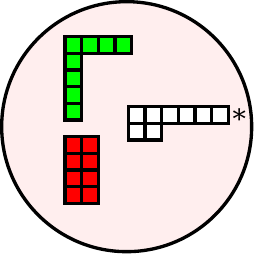}\end{matrix}, \, \begin{matrix} \includegraphics[width=64pt]{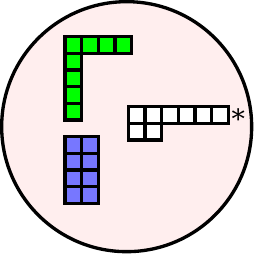}\end{matrix}, \, \begin{matrix} \includegraphics[width=64pt]{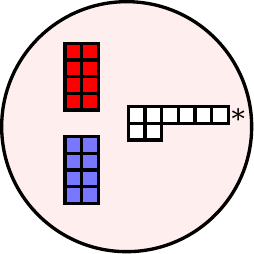}\end{matrix}$&0&none\\
\hline
$\begin{gathered}\begin{matrix} \includegraphics[width=64pt]{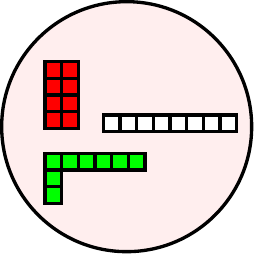}\end{matrix}\, ,\, \begin{matrix} \includegraphics[width=64pt]{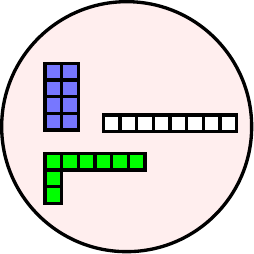}\end{matrix}\, ,\, \begin{matrix} \includegraphics[width=64pt]{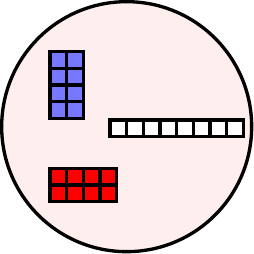}\end{matrix}\, ,\\ \begin{matrix} \includegraphics[width=64pt]{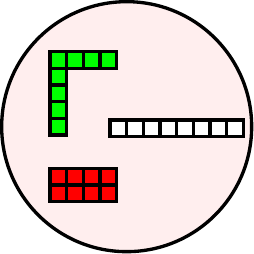}\end{matrix}\, ,\, \begin{matrix} \includegraphics[width=64pt]{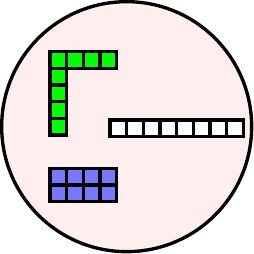}\end{matrix}\, ,\, \begin{matrix} \includegraphics[width=64pt]{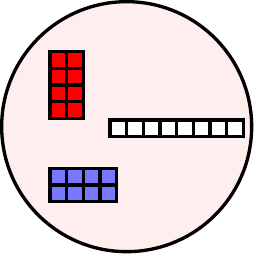}\end{matrix}\end{gathered}$&24&\mbox{\shortstack{$\tfrac{1}{2}(1,4,8_u)+\tfrac{1}{2}(2,1,8_d)$,\\ where $8_{u/d} = 8_{v},\, 8_s,\, \text{or}\, 8_c$\\ depending on whether the\\ upper/lower left-hand\\ puncture is coloured\\ green, red, or blue.}}\\
\hline
$\begin{gathered}\begin{matrix} \includegraphics[width=64pt]{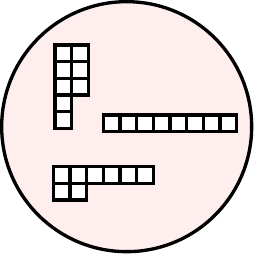}\end{matrix}\end{gathered}$&24&$\begin{aligned}\tfrac{1}{2}(2,1,1,8_v)\\ +\tfrac{1}{2}(1,2,1,8_s)\\ +\tfrac{1}{2}(1,1,2,8_c)\end{aligned}$\\
\hline$\begin{gathered}\begin{matrix} \includegraphics[width=64pt]{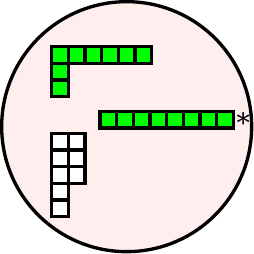}\end{matrix}\, , \,\begin{matrix} \includegraphics[width=64pt]{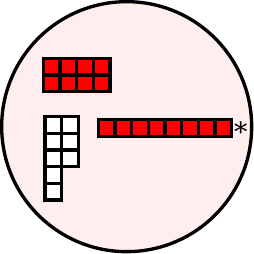}\end{matrix}\, , \,\begin{matrix} \includegraphics[width=64pt]{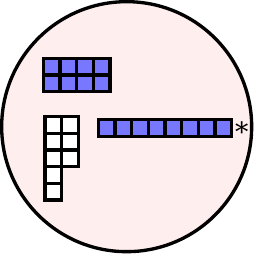}\end{matrix}\end{gathered}$&16&$\tfrac{1}{2}(4,8)$\\
\hline
$\begin{gathered}\begin{matrix} \includegraphics[width=64pt]{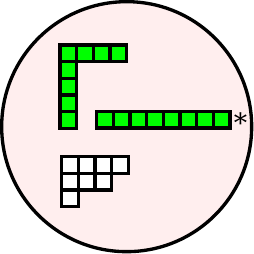}\end{matrix}\, , \, \begin{matrix} \includegraphics[width=64pt]{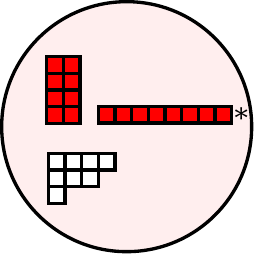}\end{matrix}\, , \, \begin{matrix} \includegraphics[width=64pt]{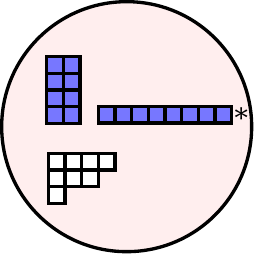}\end{matrix}\end{gathered}$&15&$\tfrac{1}{2}(2,1,8)+\tfrac{1}{2}(1,2,7)$\\
\hline
\end{tabular}

\noindent
\begin{tabular}{|c|c|c|}
\hline 
Fixture&\mbox{\shortstack{\\ Number\\ of Hypers}}&Representation\\
\hline
$\begin{gathered}\begin{matrix} \includegraphics[width=64pt]{D4fig82}\end{matrix}\, , \, \begin{matrix} \includegraphics[width=64pt]{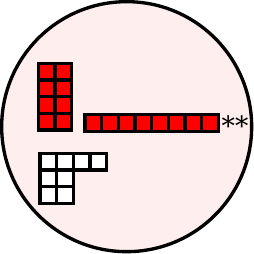}\end{matrix}\, , \, \begin{matrix} \includegraphics[width=64pt]{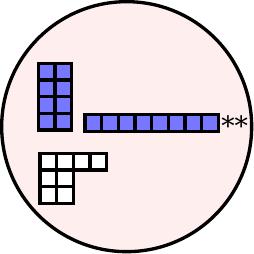}\end{matrix}\end{gathered}$&8&$(2,4)$\\
\hline
$\begin{matrix} \includegraphics[width=64pt]{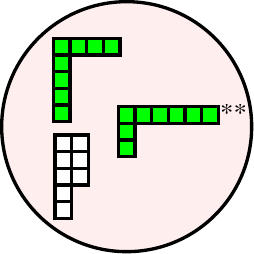}\end{matrix}\, , \,\begin{matrix} \includegraphics[width=64pt]{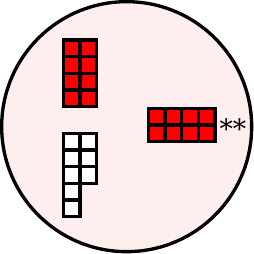}\end{matrix}\, , \,\begin{matrix} \includegraphics[width=64pt]{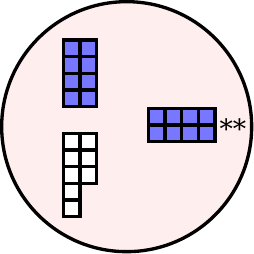}\end{matrix}$&0&none\\
\hline
$\begin{matrix} \includegraphics[width=64pt]{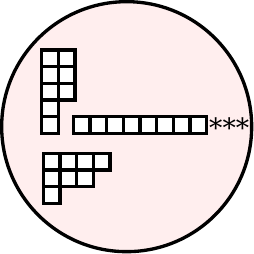}\end{matrix}$&7&$\tfrac{1}{2}(2,7)$\\
\hline
$\begin{matrix} \includegraphics[width=64pt]{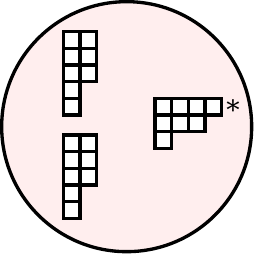}\end{matrix}$&1&$\tfrac{1}{2}(2)$\\
\hline
$\begin{matrix} \includegraphics[width=64pt]{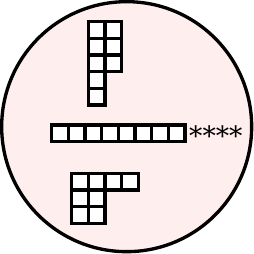}\end{matrix}$&0&none\\
\hline
\end{tabular}
\medskip

Note that, among the free field fixtures, are six which are empty (zero hypermultiplets). It might, at first blush, seem peculiar to assign global symmetry groups (${SU(2)}_8^2$ and ${SU(2)}_8$, respectively) to the regular punctures on them. However, they are attached to the rest of the surface by an $SU(2)$ cylinder, which gauges an $SU(2)$ subgroup of the global symmetry group of the attaching puncture. The centralizer of that $SU(2)$ is, respectively ${SU(2)}_8^2$ or ${SU(2)}_8$. That centralizer is what is detected by the punctures on the ostensibly ``{}empty''{} fixture. Similar remarks applied to the analogous fixtures that we saw in the $D_3$ and $A_{N-1}$ cases, studied in \cite{Chacaltana:2010ks}.

\subsubsection{{Interacting fixtures}}\label{D4interacting}

Interacting fixtures are those that contain a non-Lagrangian SCFT (e.g., the Minahan-Nemeschansky $E_n$ theories \cite{Minahan:1996cj}), and \emph{no} accompanying free hypermultiplets.

{\parindent=-0.3in
\begin{tabular}{|c|c|c|c|l|}
\hline
Fixture&\small$(d_2,d_3,d_4,d_5,d_6)$&$(a,c)$&${(G_{\text{global}})}_k$&Theory\\
\hline 
$\begin{matrix} \includegraphics[width=58pt]{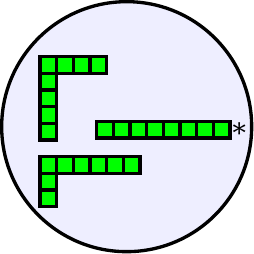}\end{matrix}\, ,\,\begin{matrix} \includegraphics[width=58pt]{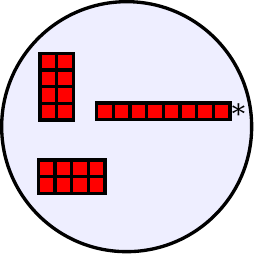}\end{matrix}\, ,\,\begin{matrix} \includegraphics[width=58pt]{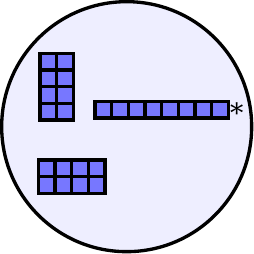}\end{matrix}$&$(0,0,1,0,0)$&$(\tfrac{59}{24},\tfrac{19}{6})$&${(E_7)}_8$&\mbox{\shortstack{The $E_7$\\ SCFT}}\\
\hline
$\begin{matrix} \includegraphics[width=58pt]{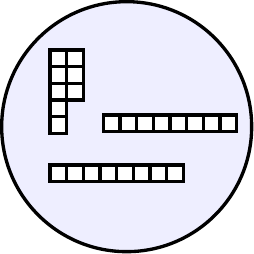}\end{matrix}$&$(0,0,0,0,1)$&$(\tfrac{95}{24},\tfrac{31}{6})$&$(E_8)_{12}$&\mbox{\shortstack{The $E_8$\\ SCFT}}\\
\hline 
$\begin{matrix} \includegraphics[width=58pt]{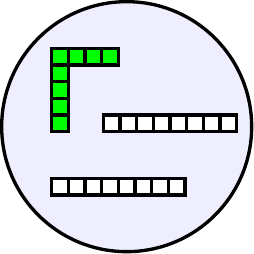}\end{matrix}\, , \,\begin{matrix} \includegraphics[width=58pt]{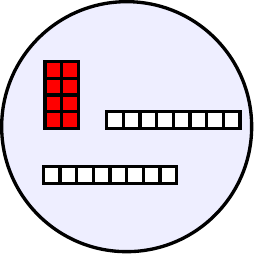}\end{matrix}\, ,\,\begin{matrix} \includegraphics[width=58pt]{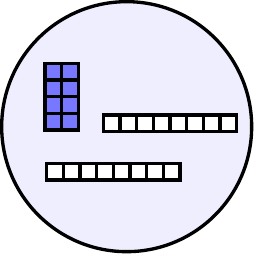}\end{matrix}$&$(0,0,1,0,1)$&$(\tfrac{23}{4},7)$&$\begin{aligned}{Spin(16)}_{12}\\ \times {SU(2)}_8\end{aligned}$&$\,$\\
\hline
$\begin{matrix} \includegraphics[width=58pt]{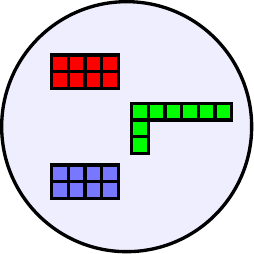}\end{matrix}$&$(0,0,1,0,1)$&$(\tfrac{65}{12},\tfrac{19}{3})$&${Sp(6)}_8$&$\,$\\
\hline 
$\begin{matrix} \includegraphics[width=58pt]{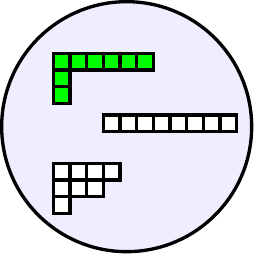}\end{matrix}\, , \,\begin{matrix} \includegraphics[width=58pt]{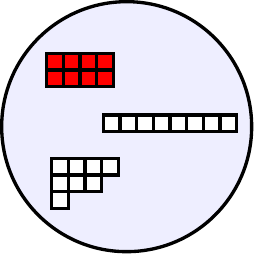}\end{matrix}\, ,\,\begin{matrix} \includegraphics[width=58pt]{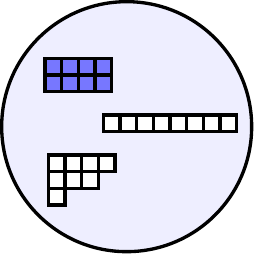}\end{matrix}$&$(0,0,1,0,2)$&$(\tfrac{25}{3},\tfrac{113}{12})$&$\begin{gathered}{Spin(9)}_{12}\\ \times{Sp(2)}_8\\ \times{SU(2)}_7\end{gathered}$&$\,$\\
\hline 
$\begin{matrix} \includegraphics[width=58pt]{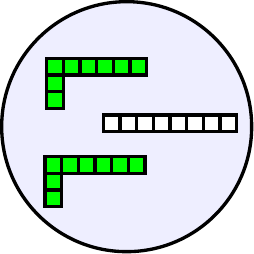}\end{matrix}\, , \,\begin{matrix} \includegraphics[width=58pt]{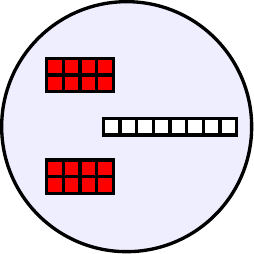}\end{matrix}\, ,\,\begin{matrix} \includegraphics[width=58pt]{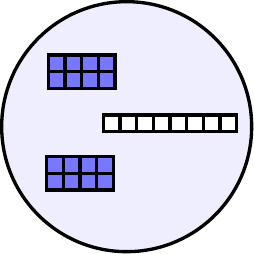}\end{matrix}$&$(0,0,2,0,2)$&$(\tfrac{61}{6},\tfrac{34}{3})$&$\begin{gathered}{Spin(9)}_{12}\\ \times {Sp(2)}^2_8\end{gathered}$&$\,$\\
\hline
$\begin{matrix} \includegraphics[width=58pt]{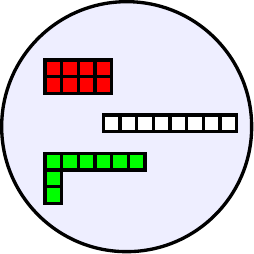}\end{matrix}\, , \,\begin{matrix} \includegraphics[width=58pt]{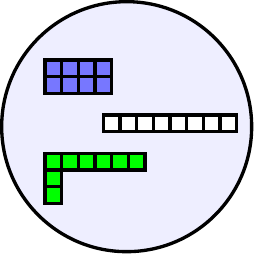}\end{matrix}\, , \,\begin{matrix} \includegraphics[width=58pt]{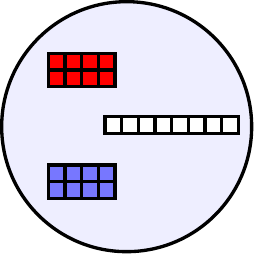}\end{matrix}$&$(0,0,2,0,2)$&$(\tfrac{61}{6},\tfrac{34}{3})$&$\begin{gathered}{Spin(8)}_{12}\\ \times {Sp(2)}^2_8\end{gathered}$&$\,$\\
\hline 
$\begin{matrix} \includegraphics[width=58pt]{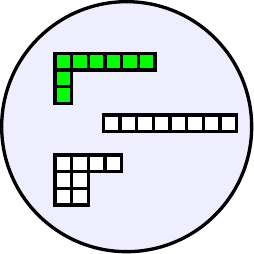}\end{matrix}\, , \,\begin{matrix} \includegraphics[width=58pt]{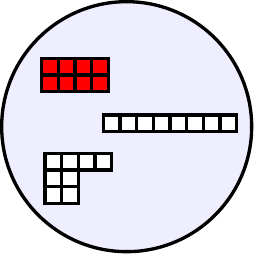}\end{matrix}\, ,\,\begin{matrix} \includegraphics[width=58pt]{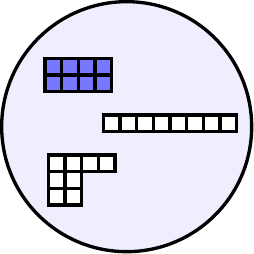}\end{matrix}$&$(0,1,1,0,1)$&$(\tfrac{163}{24},\tfrac{47}{6})$&$\begin{gathered}{Spin(10)}_{12}\\ \times{Sp(2)}_8\\ \times {U(1)}\end{gathered}$&$\,$\\
\hline
$\begin{matrix} \includegraphics[width=58pt]{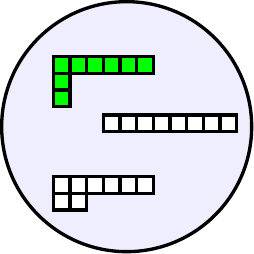}\end{matrix}\, , \,\begin{matrix} \includegraphics[width=58pt]{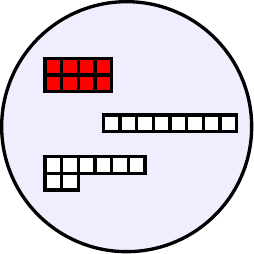}\end{matrix}\, ,\,\begin{matrix} \includegraphics[width=58pt]{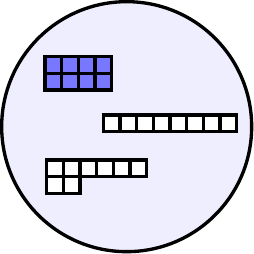}\end{matrix}$&$(0,0,3,0,2)$&$(\tfrac{287}{24},\tfrac{79}{6})$&$\begin{gathered}{Spin(8)}_{12}\\ \times{Sp(2)}_8\\ \times{SU(2)}_8^3\end{gathered}$&$\,$\\
\hline
$\begin{matrix} \includegraphics[width=58pt]{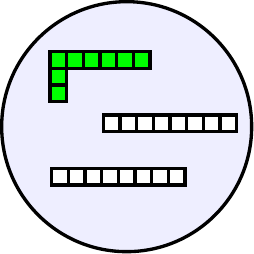}\end{matrix}\, , \,\begin{matrix} \includegraphics[width=58pt]{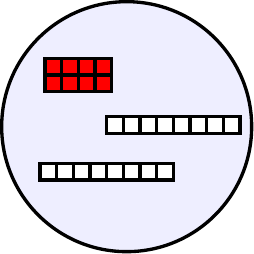}\end{matrix}\, , \,\begin{matrix} \includegraphics[width=58pt]{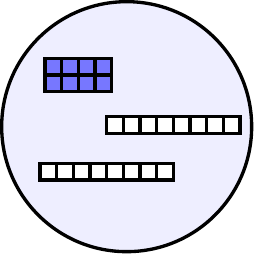}\end{matrix}$&$(0,0,3,0,3)$&$(\tfrac{179}{12},\tfrac{49}{3})$&$\begin{gathered}{Spin(8)}_{12}^2\\ \times{Sp(2)}_8\end{gathered}$&$\,$\\
\hline
\end{tabular}

\noindent
\begin{tabular}{|c|c|c|c|l|}
\hline
Fixture&\small$(d_2,d_3,d_4,d_5,d_6)$&$(a,c)$&${(G_{\text{global}})}_k$&Theory\\
\hline 
$\begin{matrix} \includegraphics[width=58pt]{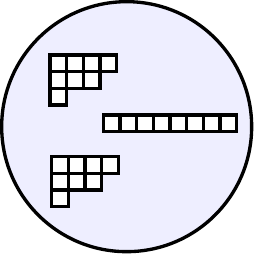}\end{matrix}$&$(0,0,0,0,2)$&$(\tfrac{13}{2},\tfrac{15}{2})$&$\begin{gathered}{(F_4)}_{12}\\ \times{SU(2)}^2_7\end{gathered}$&$\,$\\
\hline
$\begin{matrix} \includegraphics[width=58pt]{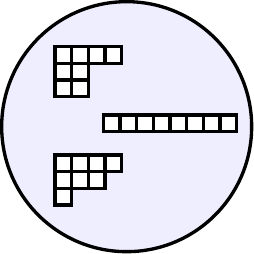}\end{matrix}$&$(0,1,0,0,1)$&$(\tfrac{119}{24},\tfrac{71}{12})$&$\begin{gathered}{(E_6)}_{12}\\ \times {SU(2)}_7\end{gathered}$&$\,$\\
\hline
$\begin{matrix} \includegraphics[width=58pt]{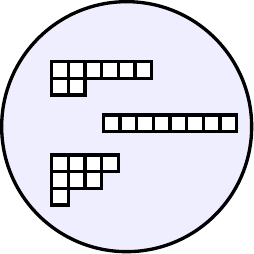}\end{matrix}$&$(0,0,2,0,2)$&$(\tfrac{81}{8},\tfrac{45}{4})$&$\begin{gathered}{Spin(8)}_{12}\\ \times {SU(2)}^3_8\\ \times {SU(2)}_7\end{gathered}$&$\,$\\
\hline 
$\begin{matrix} \includegraphics[width=58pt]{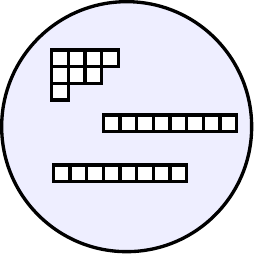}\end{matrix}$&$(0,0,2,0,3)$&$(\tfrac{157}{12},\tfrac{173}{12})$&$\begin{gathered}{Spin(8)}^2_{12}\\ \times{SU(2)}_7\end{gathered}$&$\,$\\
\hline
$\begin{matrix} \includegraphics[width=58pt]{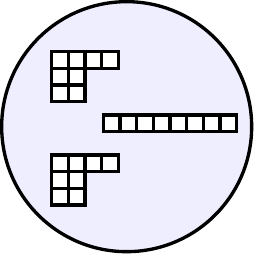}\end{matrix}$&$(0,2,0,0,0)$&$(\tfrac{41}{12},\tfrac{13}{3})$&${(E_6)}_{6}^2$&\mbox{\shortstack{Two copies\\ of the\\ $E_6$ SCFT}}\\
\hline
$\begin{matrix} \includegraphics[width=58pt]{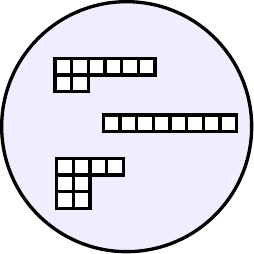}\end{matrix}$&$(0,1,2,0,1)$&$(\tfrac{103}{12},\tfrac{29}{3})$&$\begin{gathered}{Spin(8)}_{12}\\ \times{SU(2)}_8^3\\ \times {U(1)}^2\end{gathered}$&$\,$\\
\hline
$\begin{matrix} \includegraphics[width=58pt]{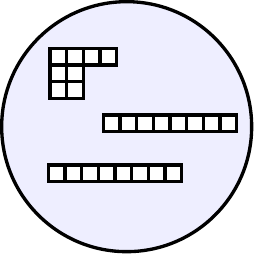}\end{matrix}$&$(0,1,2,0,2)$&$(\tfrac{277}{24},\tfrac{77}{6})$&$\begin{gathered}{Spin(8)}_{12}^2\\ \times {U(1)}^2\end{gathered}$&$\,$\\
\hline
$\begin{matrix} \includegraphics[width=58pt]{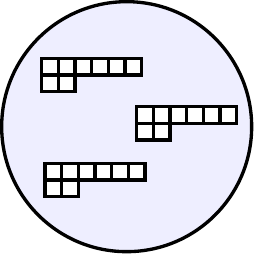}\end{matrix}$&$(0,0,4,0,1)$&$(\tfrac{259}{24},\tfrac{71}{6})$&${SU(2)}^9_8$&$\,$\\
\hline
$\begin{matrix} \includegraphics[width=58pt]{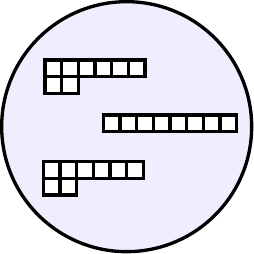}\end{matrix}$&$(0,0,4,0,2)$&$(\tfrac{55}{4},15)$&$\begin{gathered}{Spin(8)}_{12}\\ \times{SU(2)}_8^6\end{gathered}$&$\,$\\
\hline
$\begin{matrix} \includegraphics[width=58pt]{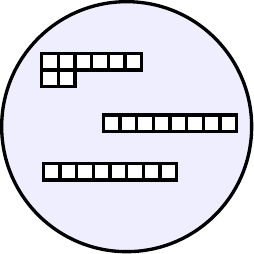}\end{matrix}$&$(0,0,4,0,3)$&$(\tfrac{401}{24},\tfrac{109}{6})$&$\begin{gathered}{Spin(8)}_{12}^2\\ \times {SU(2)}_8^3\end{gathered}$&$\,$\\
\hline
\end{tabular}

\noindent
\begin{tabular}{|c|c|c|c|l|}
\hline
Fixture&\small$(d_2,d_3,d_4,d_5,d_6)$&$(a,c)$&${(G_{\text{global}})}_k$&Theory\\
\hline
$\begin{matrix} \includegraphics[width=58pt]{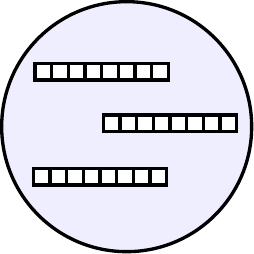}\end{matrix}$&$(0,0,4,0,4)$&$(\tfrac{59}{3},\tfrac{64}{3})$&${Spin(8)}_{12}^3$&$\,$\\
\hline
$\begin{matrix} \includegraphics[width=58pt]{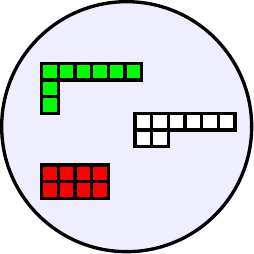}\end{matrix}\, , \,\begin{matrix} \includegraphics[width=58pt]{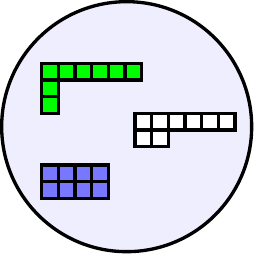}\end{matrix}\, , \,\begin{matrix} \includegraphics[width=58pt]{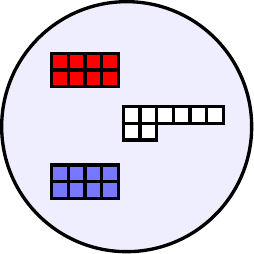}\end{matrix}$&$(0,0,2,0,1)$&$(\tfrac{173}{24},\tfrac{49}{6})$&$\begin{gathered}{Sp(3)}_8^2\\ \times {SU(2)}_8\end{gathered}$&$\,$\\
\hline
$\begin{matrix} \includegraphics[width=58pt]{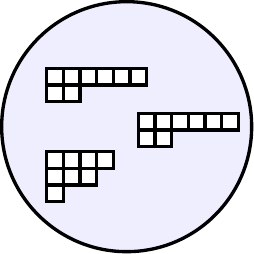}\end{matrix}$&$(0,0,2,0,1)$&$(\tfrac{43}{6},\tfrac{97}{12})$&$\begin{gathered}{Sp(2)}^3_8\\ \times {SU(2)}_7\end{gathered}$&$\,$\\
\hline
$\begin{matrix} \includegraphics[width=58pt]{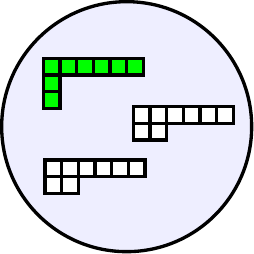}\end{matrix}\, , \,\begin{matrix} \includegraphics[width=58pt]{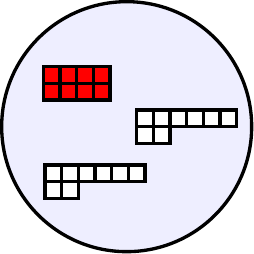}\end{matrix}\, ,\,\begin{matrix} \includegraphics[width=58pt]{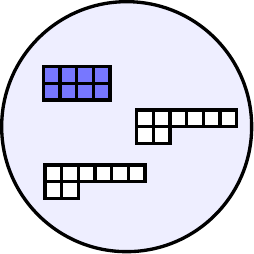}\end{matrix}$&$(0,0,3,0,1)$&$(9,10)$&$\begin{gathered}{Sp(2)}_8^2\\ \times {SU(2)}_8^4\end{gathered}$&$\,$\\
\hline
$\begin{matrix} \includegraphics[width=58pt]{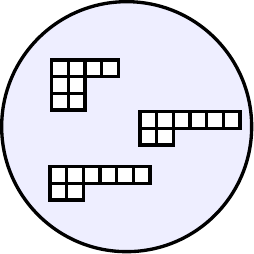}\end{matrix}$&$(0,1,2,0,0)$&$(\tfrac{45}{8},\tfrac{13}{2})$&$SU(4)^3_8$&$T_4$\\
\hline
\end{tabular}
}

\subsubsection{{Mixed fixtures}}\label{D4mixed}

Mixed fixtures are those that include an interacting SCFT, plus a number of free hypermultiplets.
\medskip

{\parindent=-0.1in{~}
\begin{tabular}{|c|c|c|c|l|}
\hline
Fixture&\small$(d_2,d_3,d_4,d_5,d_6)$&$(a,c)$&SCFT&\# Free hypers\\
\hline 
$\begin{matrix} \includegraphics[width=58pt]{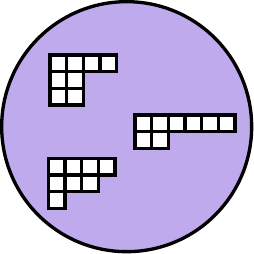}\end{matrix}$&$(0,1,0,0,0)$&$(2,\tfrac{11}{4})$&$(E_6)_6$&\mbox{\shortstack{$7$ hypers,\\ transforming as\\ $\begin{gathered}\tfrac{1}{2}(2;1,1,1)\\ +(1;2,1,1)\\ +(1;1,2,1)\\ +(1;1,1,2)\end{gathered}$}}\\
\hline 
$\begin{matrix} \includegraphics[width=58pt]{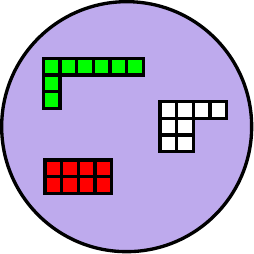}\end{matrix}\, , \,\begin{matrix} \includegraphics[width=58pt]{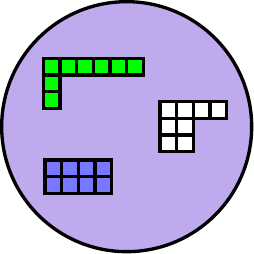}\end{matrix}\, , \,\begin{matrix} \includegraphics[width=58pt]{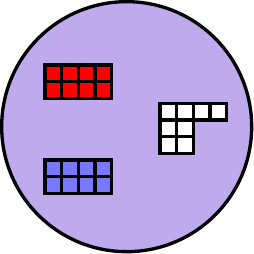}\end{matrix}$&$(0,1,0,0,0)$&$(\tfrac{49}{24},\tfrac{17}{6})$&$(E_6)_6$&\mbox{\shortstack{$8$ hypers,\\ transforming as\\ $(4;1)+(1;4)$}}\\
\hline 
\end{tabular}
}

{\parindent=-0.4in{~}
\begin{tabular}{|c|c|c|c|l|}
\hline
Fixture&\small$(d_2,d_3,d_4,d_5,d_6)$&$(a,c)$&SCFT&\# Free hypers\\
\hline 
$\begin{matrix} \includegraphics[width=58pt]{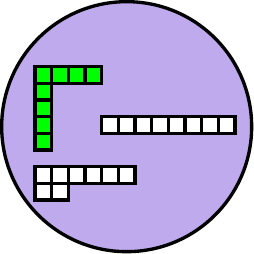}\end{matrix}\, ,\,\begin{matrix} \includegraphics[width=58pt]{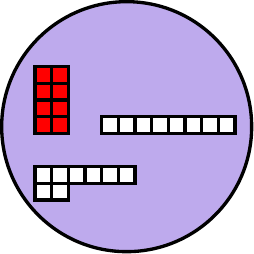}\end{matrix}\, ,\,\begin{matrix} \includegraphics[width=58pt]{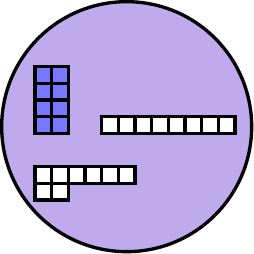}\end{matrix}$&$(0,0,1,0,0)$&$(\tfrac{67}{24},\tfrac{23}{6})$&${(E_7)}_8$&\raisebox{-7ex}{\shortstack{8 hypers,\\ transforming as\\ $\tfrac{1}{2}(1;2,1,1;8_v)$,\\ $\tfrac{1}{2}(1;1,2,1;8_s)$\\or $\tfrac{1}{2}(1;1,1,2;8_c)$,\\ depending on\\ the colour of the\\ green/red/blue\\ puncture}}\\
\hline
$\begin{matrix} \includegraphics[width=58pt]{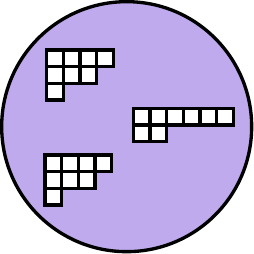}\end{matrix}$&$(0,0,0,0,1)$&$(\tfrac{85}{24},\tfrac{13}{3})$&$Sp(5)_7$&\mbox{\shortstack{3 hypers,\\ transforming as\\ $\begin{gathered}\tfrac{1}{2}(1; 1; 2, 1, 1)\\ +\tfrac{1}{2}(1; 1; 1, 2, 1)\\ +\tfrac{1}{2}(1; 1; 1, 1, 2)\end{gathered}$}}\\
\hline
$\begin{matrix} \includegraphics[width=58pt]{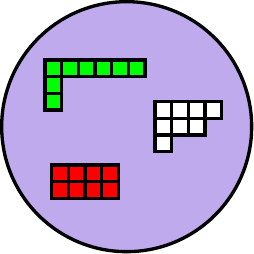}\end{matrix}\, , \,\begin{matrix} \includegraphics[width=58pt]{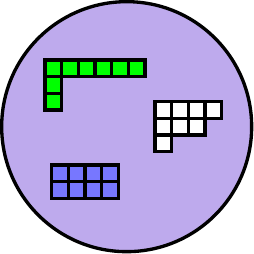}\end{matrix}\, , \,\begin{matrix} \includegraphics[width=58pt]{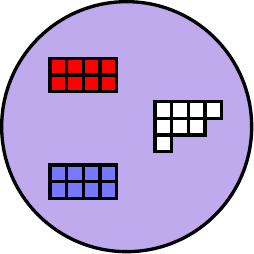}\end{matrix}$&$(0,0,0,0,1)$&$(\tfrac{43}{12},\tfrac{53}{12})$&${Sp(5)}_7$&\mbox{\shortstack{4 hypers,\\ transforming as\\ $\begin{gathered}\tfrac{1}{2}(1;4;1)\\+\tfrac{1}{2}(4;1;1)\end{gathered}$}}\\
\hline
$\begin{matrix} \includegraphics[width=58pt]{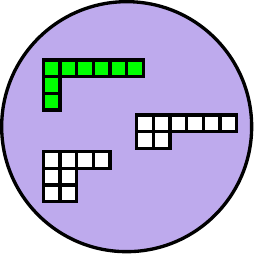}\end{matrix}\, , \,\begin{matrix} \includegraphics[width=58pt]{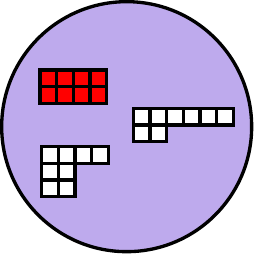}\end{matrix}\, , \,\begin{matrix} \includegraphics[width=58pt]{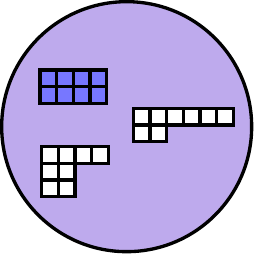}\end{matrix}$&$(0,1,1,0,0)$&$(\tfrac{23}{6},\tfrac{14}{3})$&$\begin{gathered}SU(2)_6\\ \times SU(8)_8\end{gathered}$&\mbox{\shortstack{$2$ hypers,\\ transforming as\\ $(1;2,1,1),$\\ $(1;1,2,1),$\\or $(1;2,1,1)$}}\\
\hline
$\begin{gathered}\begin{matrix} \includegraphics[width=58pt]{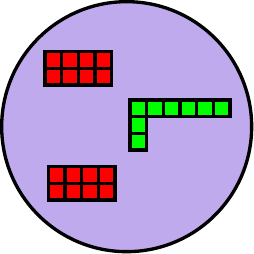}\end{matrix}\, , \, \begin{matrix} \includegraphics[width=58pt]{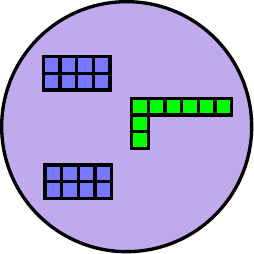}\end{matrix}\, ,\, \begin{matrix} \includegraphics[width=58pt]{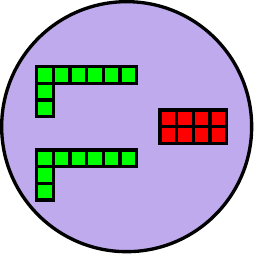}\end{matrix}\, , \\ \begin{matrix} \includegraphics[width=58pt]{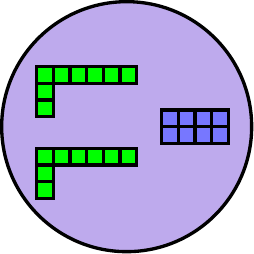}\end{matrix}\, ,\, \begin{matrix} \includegraphics[width=58pt]{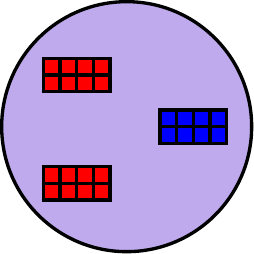}\end{matrix}\, ,\, \begin{matrix} \includegraphics[width=58pt]{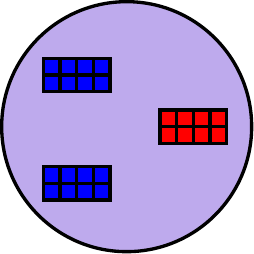}\end{matrix}\end{gathered}$&$(0,0,1,0,1)$&$(\tfrac{65}{12},\tfrac{19}{3})$&$\begin{gathered}{Sp(4)}_8\\ \times {Sp(2)}_7\end{gathered}$&\mbox{\shortstack{2 hypers,\\ transforming as\\ $\tfrac{1}{2}(1;1;4)$}}\\
\hline
$\begin{matrix} \includegraphics[width=58pt]{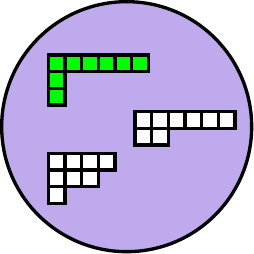}\end{matrix}\, , \,\begin{matrix} \includegraphics[width=58pt]{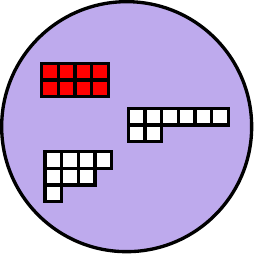}\end{matrix}\, , \,\begin{matrix} \includegraphics[width=58pt]{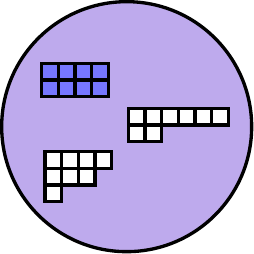}\end{matrix}$&$(0,0,1,0,1)$&$(\tfrac{43}{8},\tfrac{25}{4})$&$\begin{gathered}{Sp(4)}_8\\ \times {Sp(2)}_7\end{gathered}$&\mbox{\shortstack{1 hyper,\\ transforming as\\ $\begin{gathered}\tfrac{1}{2}(1;1;2,1,1),\\ \tfrac{1}{2}(1;1;1,2,1)\\ \text{or}\, \tfrac{1}{2}(1;1;1,1,2)\end{gathered}$}}\\
\hline
$\begin{matrix} \includegraphics[width=58pt]{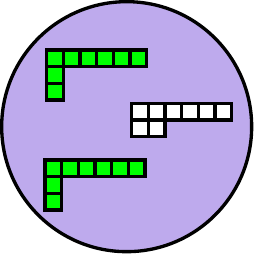}\end{matrix}\, , \,\begin{matrix} \includegraphics[width=58pt]{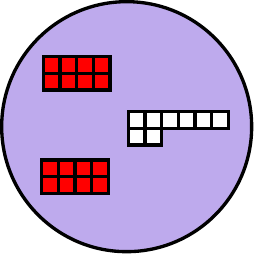}\end{matrix}\, , \,\begin{matrix} \includegraphics[width=58pt]{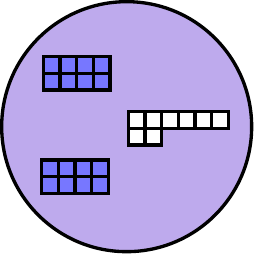}\end{matrix}$&$(0,0,2,0,1)$&$(\tfrac{173}{24},\tfrac{49}{6})$&$\begin{gathered}{Sp(2)}_8^3\\ \times {SU(2)}_7\end{gathered}$&\mbox{\shortstack{1 hyper,\\ transforming as\\ $\begin{gathered}\tfrac{1}{2}(1;1;2,1,1),\\ \tfrac{1}{2}(1;1;1,2,1)\\ \text{or}\, \tfrac{1}{2}(1;1;1,1,2)\end{gathered}$}}\\
\hline
\end{tabular}
}
\subsection{{The ${Sp(4)}_8\times {Sp(2)}_7$ and ${Sp(5)}_7$ SCFTs}}\label{D4anomSCFTs}

A couple of SCFTs make a somewhat unusual appearance in the above list of mixed fixtures. Usually, the mixed fixtures contain SCFTs which have previously appeared elsewhere (without the additional hypermultiplets). Indeed, ${(E_6)}_6$, ${(E_7)}_8$ and ${SU(2)}_6\times{SU(8)}_8$ SCFTs (the latter was called the ``$R_{0,4}$ theory" in \cite{Chacaltana:2010ks}) have all appeared previously. In the present case, we find two new ones, which do not appear to arise \emph{in the absence} of accompanying hypermultiplets.

\subsubsection{${Sp(4)}_8\times {Sp(2)}_7$ SCFT}\label{Sp4Sp2}

One is the ${Sp(4)}_8\times {Sp(2)}_7$ SCFT. It has $(a,c)=\left(\tfrac{16}{3},\tfrac{37}{6}\right)$, and graded Coulomb branch dimension $(d_2,d_3,d_4,d_5,d_6)=(0,0,1,0,1)$. Its global symmetry group is

\begin{displaymath}
G_X = {Sp(4)}_8\times {Sp(2)}_7
\end{displaymath}
It appears in our table, accompanied by either 1 hypermultiplet (3 fixtures) or 2 hypermultiplets (6 fixtures).

Let us look a couple of examples of its appearance.

Consider a $Spin(7)$ gauge theory, with matter in the $3(8)+2(7)+1$.

\begin{displaymath}
\begin{aligned}
\begin{matrix} \includegraphics[width=279pt]{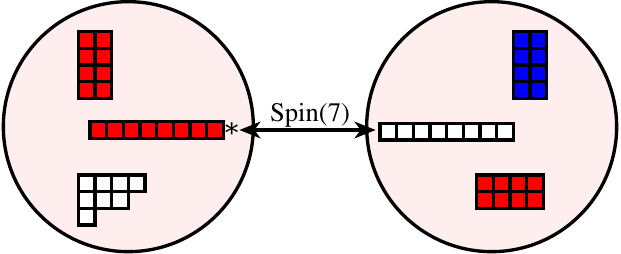}\end{matrix}
\end{aligned}
\end{displaymath}
This theory has two distinct strong-coupling points. One,

\begin{displaymath}
\begin{aligned}
\begin{matrix} \includegraphics[width=279pt]{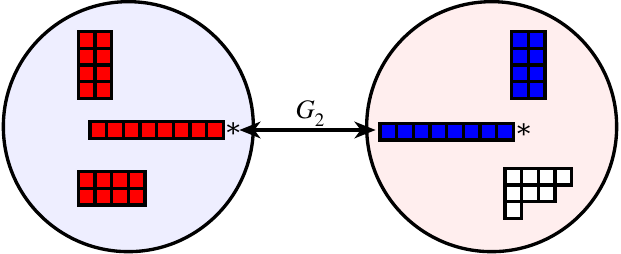}\end{matrix}
\end{aligned}
\end{displaymath}
is a $G_2$ gauge theory, with matter in the $2(7)+1$, coupled to the ${(E_7)}_8$ SCFT. Aside from the addition of the free hypermultiplet, this was example 10 of Argyres and Wittig \cite{Argyres:2007tq}.

The other strong coupling point of this theory,

\begin{displaymath}
\begin{aligned}
\begin{matrix} \includegraphics[width=279pt]{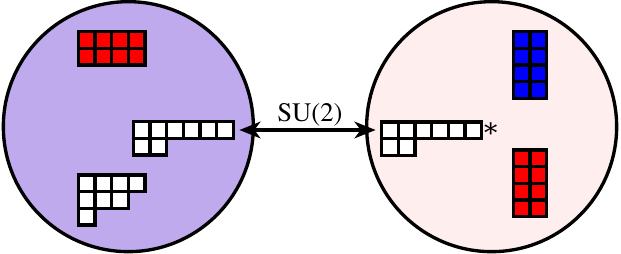}\end{matrix}
\end{aligned}
\end{displaymath}
is an $SU(2)$ gauge theory coupled to the ${Sp(4)}_8\times {Sp(2)}_7$ SCFT. The fixture on the right is empty; the mixed-fixture on the left provides both the SCFT and an additional free hypermultiplet.

As a second example, consider

\begin{displaymath}
\begin{aligned}
\begin{matrix} \includegraphics[width=279pt]{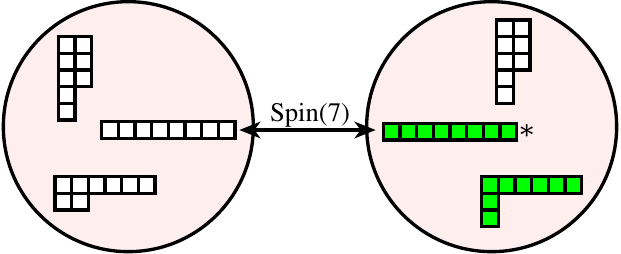}\end{matrix}
\end{aligned}
\end{displaymath}
This is a $Spin(7)$ gauge theory, with matter in the $4(8)+(7)+(1)$. The S-dual theory

\begin{displaymath}
\begin{aligned}
\begin{matrix} \includegraphics[width=279pt]{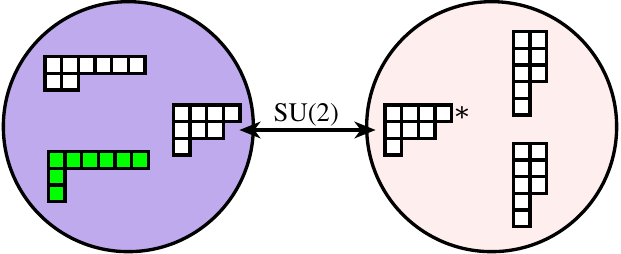}\end{matrix}
\end{aligned}
\end{displaymath}
is an $SU(2)$ gauge theory. The fixture on the right contributes a half-hypermultiplet in the fundamental. The fixture on the left is the ${Sp(4)}_8\times {Sp(2)}_7$ SCFT plus a single \emph{free} hypermultiplet. We weakly gauge an $SU(2)$ subgroup of ${Sp(2)}_7\subset G_X$. From both points of view, we reproduce

\begin{displaymath}
G_{\text{global}} = Sp(4)_8\times SU(2)_7 + 1\, \text{free hypermultiplet}
\end{displaymath}
A third example is provided by the S-dual of $Spin(8)$ gauge theory with matter in the $4(8_s)+2(8_c)$. This is discussed in section \S\ref{Spin8-42}.

\subsubsection{${Sp(5)}_7$ SCFT}\label{Sp5}

The other ``{}new''{} SCFT is the ${Sp(5)}_7$ SCFT. It has $(a,c)= \left(\tfrac{41}{12},\tfrac{49}{12}\right)$ and a Coulomb branch of graded dimension $(d_2,\dots,d_6)= (0,0,0,0,1)$. The global symmetry group is ${Sp(5)}_7$.

The ${Sp(5)}_7$ SCFT appears twice on our list, once accompanied accompanied by 3 hypermultiplets (transforming as the $\tfrac{1}{2}(1; 1; 2, 1, 1)+\tfrac{1}{2}(1; 1; 1, 2, 1)+\tfrac{1}{2}(1; 1; 1, 1, 2)$ of the manifest $SU(2)\times SU(2)\times SU(2)^3$ associated to the punctures), and once (3 fixtures) accompanied by 4 hypermultiplets (transforming as the $\tfrac{1}{2}(1;4;1)+\tfrac{1}{2}(4;1;1)$ of the manifest $Sp(2)\times Sp(2)\times SU(2)$ associated to the punctures).

Let us look at some examples of the ${Sp(5)}_7$ SCFT. Consider the 4-punctured sphere

\begin{displaymath}
\begin{aligned}
\begin{matrix} \includegraphics[width=279pt]{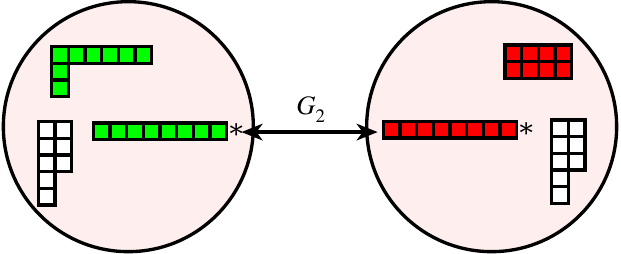}\end{matrix}
\end{aligned}
\end{displaymath}
Both fixtures provide 2 hypers in the 7 of $G_2$, plus 2 free hypers, so the 4-punctured sphere represents the $G_2$ theory with 4 hypers in the 7, plus 4 free hypers.

\begin{displaymath}
G_{\text{global}} = {Sp(4)}_7 + \text{4 free hypers}
\end{displaymath}
Aside from the 4 free hypers, this is example 4 of Argyres-Wittig \cite{Argyres:2007tq}.

The S-dual theory is

\begin{displaymath}
\begin{aligned}
\begin{matrix} \includegraphics[width=279pt]{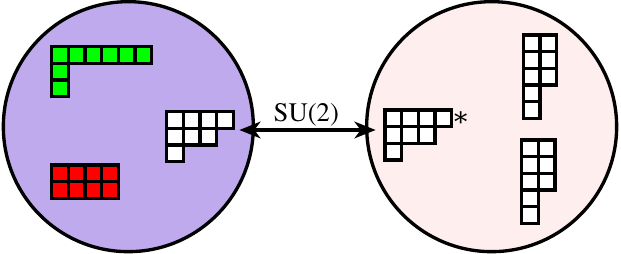}\end{matrix}
\end{aligned}
\end{displaymath}
The fixture on the left is the ${Sp(5)}_7$ SCFT, with 4 free hypers. The fixture on the right contributes a half-hyper in the fundamental of $SU(2)$. Gauging an $SU(2)\subset {Sp(5)}_7$, yields the expected ${Sp(4)}_7$ global symmetry group of the S-dual of $G_2$ with 4 fundamentals.

As another example, consider the 4-punctured sphere

\begin{displaymath}
\begin{aligned}
\begin{matrix} \includegraphics[width=279pt]{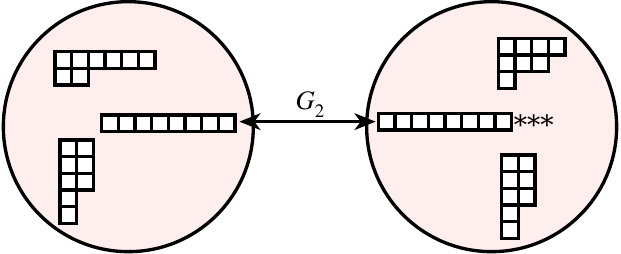}\end{matrix}
\end{aligned}
\end{displaymath}
Here the fixture on the left represents 3 hypers in the 7 of $G_2$ plus 3 free hypers, and the fixture on the right represents 1 hyper in the 7. Notice that the $G_2$ cylinder in this example is different from the one in the previous example.

S-dualizing, we obtain

\begin{displaymath}
\begin{aligned}
\begin{matrix} \includegraphics[width=279pt]{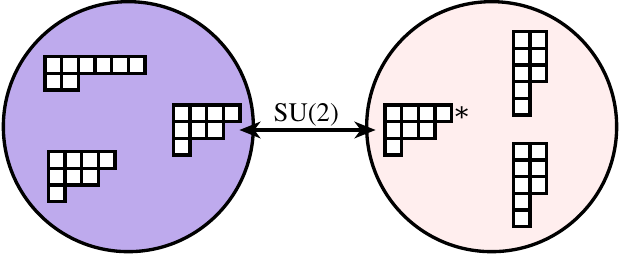}\end{matrix}
\end{aligned}
\end{displaymath}
The fixture on the left is the ${Sp(5)}_7$ SCFT, where we gauge an $SU(2)\subset Sp(5)$, accompanied by 3 free hypers. The fixture on the right contributes 1 fundamental half-hyper.

A third example, also involving $G_2$, is

\begin{displaymath}
\begin{aligned}
\begin{matrix} \includegraphics[width=279pt]{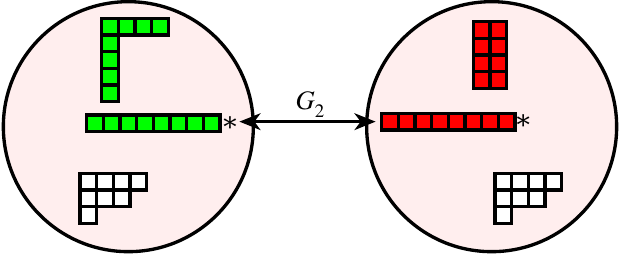}\end{matrix}
\end{aligned}
\end{displaymath}
This is $G_2$ with 4 fundamentals and \emph{two} free hypermultiplets.

The S-dual is

\begin{displaymath}
\begin{aligned}
\begin{matrix} \includegraphics[width=279pt]{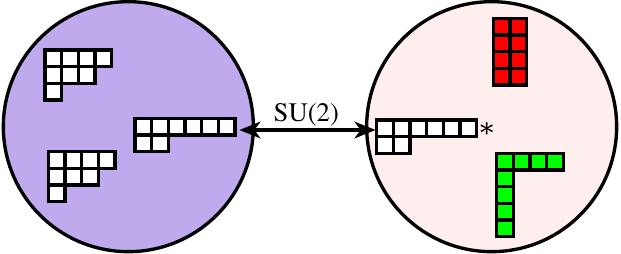}\end{matrix}
\end{aligned}
\end{displaymath}
The fixture on the right is empty. The fixture on the left is, again the ${Sp(5)}_7$ SCFT, with one hypermultiplet transforming as a half-hyper in the fundamental of $SU(2)$ and two free hypermultiplets.

For a non-$G_2$-related example, consider

\begin{displaymath}
\begin{aligned}
\begin{matrix} \includegraphics[width=279pt]{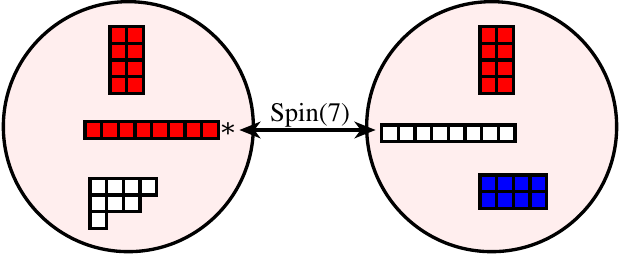}\end{matrix}
\end{aligned}
\end{displaymath}
The fixture on the left contributes hypermultiplets in the $2(7) + 1$. The fixture on the right is an $8_s+2(8_c)$, considered as a representation of $Spin(8)$. Under the chosen embedding of $Spin(7)$, the $8_s$ decomposes as $7+1$, which the $8_c$ (and also the $8_v$) decomposes as the $8$. So, all-in-all, this is a $Spin(7)$ gauge theory, with matter in the $3(7)+2(8) +2(1)$, so

\begin{displaymath}
G_{\text{global}} = {Sp(3)}_7\times {Sp(2)}_8 + \text{2 free hypers}
\end{displaymath}
The S-dual theory is

\begin{displaymath}
\begin{aligned}
\begin{matrix} \includegraphics[width=279pt]{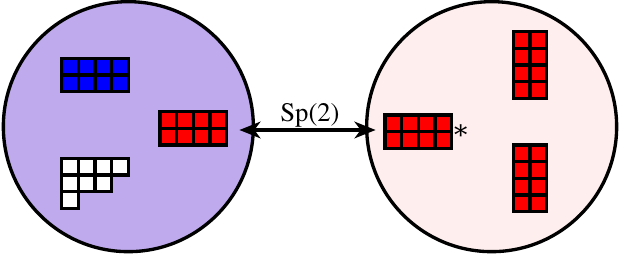}\end{matrix}
\end{aligned}
\end{displaymath}
The fixture on the right contribute 2 hypermultiplets in the fundamental of $Sp(2)$. The fixture on the left is the ${Sp(5)}_7$ SCFT, accompanied by 4 hypermultiplets, two of which form an additional half-hypermultiplet in the fundamental of $Sp(2)$ and two of which are free. Altogether, there are 5 half-hypermultiplets in the fundamental, yielding the $Spin(5)= {Sp(2)}_8$ factor in $G_{\text{global}}$. Gauging the $Sp(2)\subset {Sp(5)}_7$ yields the remaining ${Sp(3)}_7$. This is example 5 of Argyres and Wittig \cite{Argyres:2007tq}.

\section{{$Spin(8)$ Gauge Theory}}\label{Spin8}

$Spin(8)$ gauge theory ---{} with $n_s$ hypermultiplets in the $8_s$, $n_c$ hypermultiplets in the $8_c$ and $n_v$ hypermultiplets in the $8_v$ ---{} has vanishing $\beta$-function for $n_s+n_c+n_v=6$. The global symmetry group is

\begin{displaymath}
G_{\text{global}}= {Sp(n_s)}_8\times{Sp(n_c)}_8\times{Sp(n_v)}_8
\end{displaymath}
In the $D_4$ theory, all of the cases, with $n_{s,c,v}\leq 4$, are realized on the 4-punctured sphere. Up to $Spin(8)$ triality, this yields five different cases. We will discuss each of them, in turn, and give the strong-coupling behaviour in each case.

For the cases of $(n_s,n_c,n_v)=(3,2,1)$ and $(3,3,0)$, Argyres and Wittig \cite{Argyres:2007tq} conjectured a strong-coupling dual. We find that each of these cases has \emph{two} distinct strong-coupling limits. In each case, the conjecture of Argyres and Wittig corresponds to one of the two strong-coupling limits, that we find.

\subsection{{$2(8_s)+2(8_c)+2(8_v)$}}\label{Spin8-222}

The dual of $Spin(8)$, with matter in the $2(8_s)+2(8_c)+2(8_v)$, is an $SU(2)$ gauge theory, coupled to a half-hypermultiplet in the fundamental, and to the ${Sp(2)}^3_8\times {SU(2)}_7$ SCFT.

One realization is

\begin{displaymath}
\begin{aligned}
\begin{matrix} \includegraphics[width=279pt]{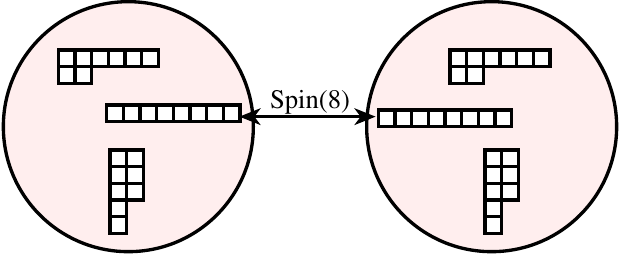}\end{matrix}
\end{aligned}
\end{displaymath}
Each fixture contributes one $(8_v+8_s+8_c)$. The S-dual theory is

\begin{displaymath}
\begin{aligned}
\begin{matrix} \includegraphics[width=279pt]{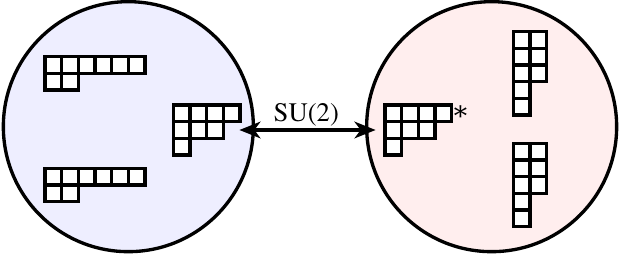}\end{matrix}
\end{aligned}
\end{displaymath}
where the fixture on the right is a half-hypermultiplet in the fundamental of $SU(2)$, and the fixture on the left is the ${Sp(2)}^3_8\times {SU(2)}_7$ SCFT.

Another realization of the same theory is

\begin{displaymath}
\begin{aligned}
\begin{matrix} \includegraphics[width=279pt]{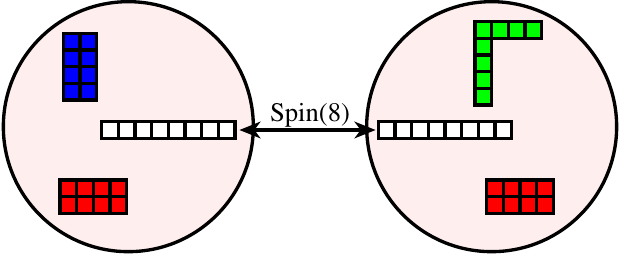}\end{matrix}
\end{aligned}
\end{displaymath}
Here, the fixture on the left contributes $8_s +2(8_c)$, and the fixture on the right contributes $8_s +2(8_v)$. The S-dual is

\begin{displaymath}
\begin{aligned}
\begin{matrix} \includegraphics[width=279pt]{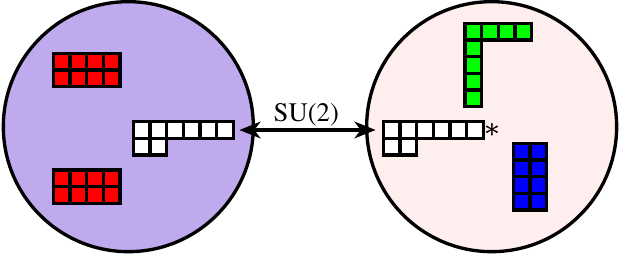}\end{matrix}
\end{aligned}
\end{displaymath}
The fixture on the right is empty; the fixture on the left is the ${Sp(2)}_8^3\times{SU(2)}_7$ SCFT plus a half-hypermultiplet in the fundamental of $SU(2)$.

\subsection{{$3(8_s)+2(8_c)+8_v$}}\label{Spin8-321}

$Spin(8)$ gauge theory, with matter in the $3(8_s)+2(8_c)+8_v$, has two \emph{distinct} strong-coupling limits. One is a $Spin(7)$ gauge theory, with matter in the $3(8)$, coupled to the ${(E_7)}_8$ SCFT. The other strong coupling limit is an $SU(2)$ gauging of the ${Sp(3)}^2_8\times {SU(2)}_8$.

One realization is

\begin{displaymath}
\begin{aligned}
\begin{matrix} \includegraphics[width=279pt]{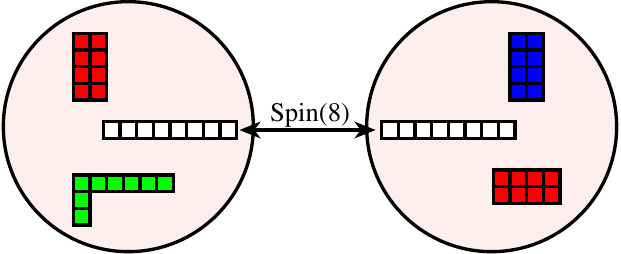}\end{matrix}
\end{aligned}
\end{displaymath}
The fixture one the left contributes $2(8_s)+8_v$, and the fixture on the right contributes $8_s +2(8_c)$.

One of the corresponding strong-coupling points is given by

\begin{displaymath}
\begin{aligned}
\begin{matrix} \includegraphics[width=279pt]{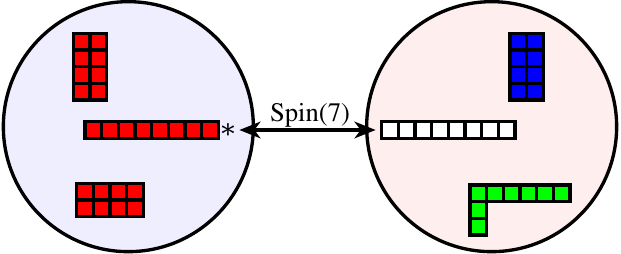}\end{matrix}
\end{aligned}
\end{displaymath}
The fixture on the right yields matter in 3 copies of the $8$; the fixture on the left is the ${(E_7)}_8$ SCFT.

The other strong coupling point is

\begin{displaymath}
\begin{aligned}
\begin{matrix} \includegraphics[width=279pt]{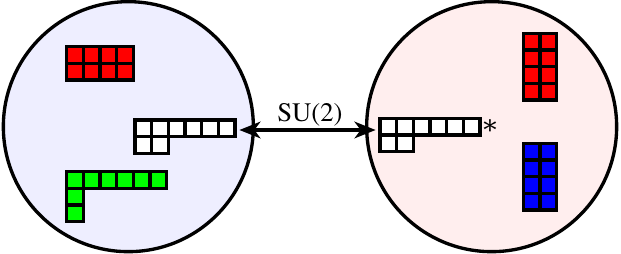}\end{matrix}
\end{aligned}
\end{displaymath}
The fixture on the right is empty, while the fixture on the left is the ${Sp(3)}_8^2\times {SU(2)}_8$ SCFT, where we gauge an $SU(2)\subset Sp(3)_8$.

Another realization of the same theory is

\begin{displaymath}
\begin{aligned}
\begin{matrix} \includegraphics[width=279pt]{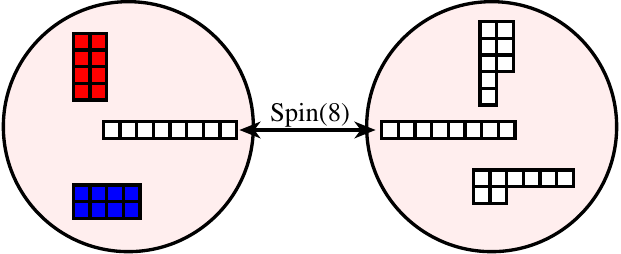}\end{matrix}
\end{aligned}
\end{displaymath}
One strong coupling point is given by

\begin{displaymath}
\begin{aligned}
\begin{matrix} \includegraphics[width=279pt]{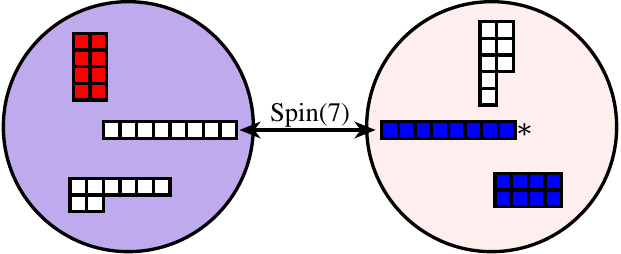}\end{matrix}
\end{aligned}
\end{displaymath}
The fixture on the right contribute 2 hypermultiplets in the $8$ of $Spin(7)$. The fixture on the left is the ${(E_7)}_8$ SCFT plus an additional hypermultiplet in the $8$.

The other strong coupling point is

\begin{displaymath}
\begin{aligned}
\begin{matrix} \includegraphics[width=279pt]{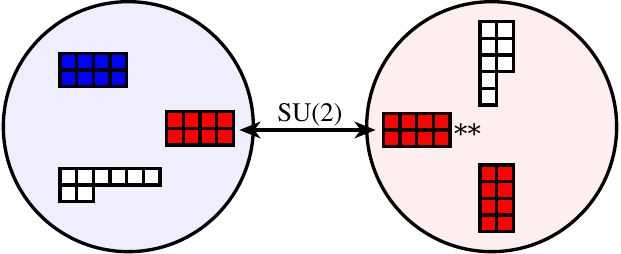}\end{matrix}
\end{aligned}
\end{displaymath}
The fixture on the right is empty; the fixture on the left is, again, the ${Sp(3)}_8^2\times {SU(2)}_8$ SCFT.

\subsection{{$3(8_s) +3(8_c)$}}\label{Spin8-33}

$Spin(8)$ gauge theory, with matter in the $3(8_s) +3(8_c)$ also has two distinct strong coupling points. One is $G_2$ gauge theory, coupled to two copies of the ${(E_7)}_8$ SCFT. The other is an $SU(2)$ gauging of the ${Sp(3)}_8^2\times {SU(2)}_8$ SCFT.

This is realized via

\begin{displaymath}
\begin{aligned}
\begin{matrix} \includegraphics[width=279pt]{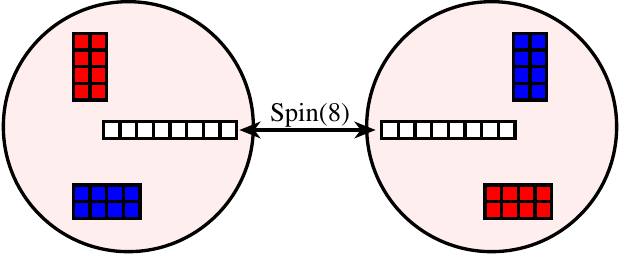}\end{matrix}
\end{aligned}
\end{displaymath}
The fixture on the left yields $2(8_s)+8_c$, while the figure on the right yields $8_S + 2(8_c)$.

One strong-coupling point is given by

\begin{displaymath}
\begin{aligned}
\begin{matrix} \includegraphics[width=279pt]{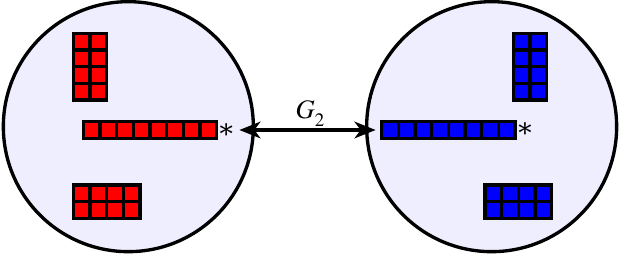}\end{matrix}
\end{aligned}
\end{displaymath}
Here, each fixture is a copy of the ${(E_7)}_8$ SCFT.

The other strong coupling point is

\begin{displaymath}
\begin{aligned}
\begin{matrix} \includegraphics[width=279pt]{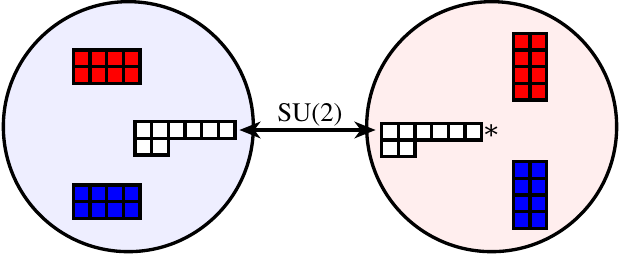}\end{matrix}
\end{aligned}
\end{displaymath}
The fixture on the right is empty. The fixture on the left is the ${Sp(3)}^2_8\times {SU(2)}_8$ SCFT where, this time, we gauge the $SU(2)_8$.

\subsection{{$4(8_s)+2(8_c)$}}\label{Spin8-42}

$Spin(8)$, with matter in the $4(8_s)+2(8_c)$ has, as its S-dual, an $Sp(2)$ gauge theory, with 5 half-hypermultiplets in the fundamental, coupled to the ${Sp(4)}_8\times {Sp(2)}_7$ SCFT.

\begin{displaymath}
\begin{aligned}
\begin{matrix} \includegraphics[width=279pt]{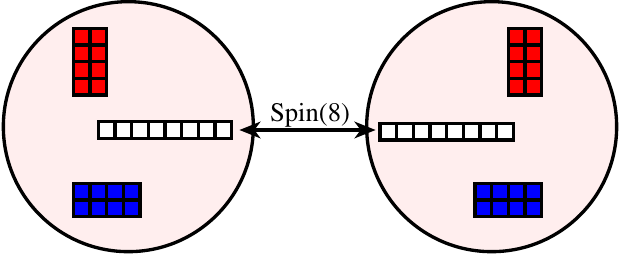}\end{matrix}
\end{aligned}
\end{displaymath}
yields a $Spin(8)$ gauge theory, with matter in the $4(8_s)+2(8_c)$.

The S-dual theory is

\begin{displaymath}
\begin{aligned}
\begin{matrix} \includegraphics[width=279pt]{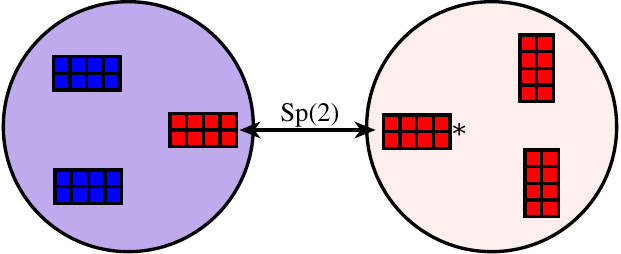}\end{matrix}
\end{aligned}
\end{displaymath}
The fixture on the right contributes two hypermultiplets in the fundamental. The fixture on the left is the ${Sp(4)}_8\times{Sp(2)}_7$ with an additional half-hypermultiplet in the fundamental of $Sp(2)$. Since there are, in total, five half-hypermultiplets in the fundamental, the flavour symmetry associated to the matter is $Spin(5)=Sp(2)_8$; the rest of $G_{\text{global}}$ comes from the $Sp(4)_8\subset {Sp(4)}_8\times{Sp(2)}_7$.

\subsection{{$4(8_s)+8_c+8_v$}}\label{Spin8-411}

Finally, $Spin(8)$ gauge theory, with matter in the $4(8_s)+8_c+8_v$ has, as its S-dual, an $Sp(2)$ gauge theory, with 2 hypermultiplets in the fundamental, coupled to the ${Sp(6)}_8$ SCFT.

The $Spin(8)$ gauge theory can be realized as

\begin{displaymath}
\begin{aligned}
\begin{matrix} \includegraphics[width=279pt]{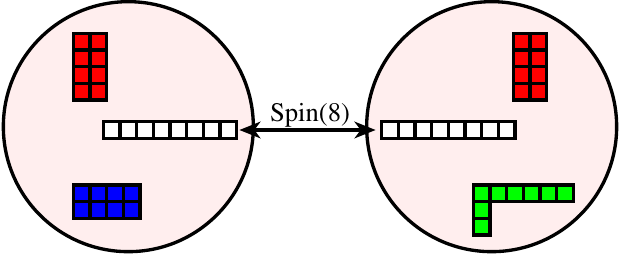}\end{matrix}
\end{aligned}
\end{displaymath}
where the fixture on the left gives matter in the $2(8_s)+8_c$ and the fixture on the right gives matter in the $2(8_s)+8_v$.

The S-dual is

\begin{displaymath}
\begin{aligned}
\begin{matrix} \includegraphics[width=279pt]{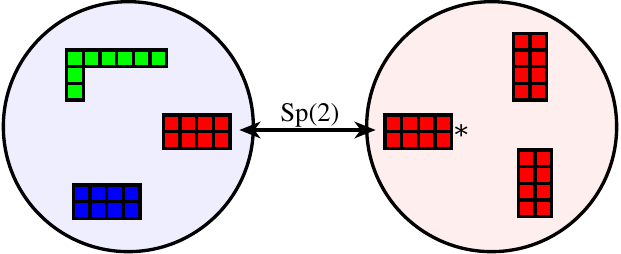}\end{matrix}
\end{aligned}
\end{displaymath}
The fixture on the right is 2 fundamental hypermultiplets of $Sp(2)$, which contribute the $Spin(4) = {SU(2)}_8^2$ factor to the global symmetry group. The fixture on the left is the ${Sp(6)}_8$ SCFT.

\subsection{Seiberg-Witten curves}\label{SWcurve}

It is straightforward to compute the Seiberg-Witten curves, associated to any of these theories, in the form \eqref{DNSW}

\begin{equation*}
0=\lambda^{8} + \sum_{k=1}^{3} \lambda^{8-2k} \phi_{2k}(y) +\tilde{\phi}^2(y)
\end{equation*}

For instance, for $Spin(8)$ gauge theory, with hypermultiplets in the $3(8_v) + 3(8_s)$, imposing the constraints, at each of the punctures, yields
\begin{equation}
\begin{split}
\phi_2(y)&= \frac{u_2\,(dy)^2}{(y-y_1)(y-y_2)(y-y_3)(y-y_4)}\\
\phi_4(y)&= \frac{[u_4\,(y-y_2)(y-y_3) - 2\tilde{u}\,(y-y_1)(y-y_4) +u_2^2\,(y-y_1)(y-y_3)/4](dy)^4}{(y-y_1)^3(y-y_2)^2(y-y_3)^3(y-y_4)^2}\\
\phi_6(y)&=\frac{[u_6\, (y-y_2)+u_2\tilde{u}\,(y_1-y_2) ](dy)^6}{(y-y_1)^4(y-y_2)^3(y-y_3)^4(y-y_4)^2}\\
\tilde{\phi}(y)&=\frac{\tilde{u}\,(dy)^4}{(y-y_1)^2(y-y_2)^2(y-y_3)^3(y-y_4)}\\
\end{split}
\end{equation}

Here $u_2,u_4,u_6$ and $\tilde{u}$ are the Coulomb branch parameters.
The obvious $SL(2,\mathbb{C})$ symmetry means that the physics depends only on the cross-ratio
$$
e(\tau) = \frac{(y_1-y_2)(y_3-y_4)}{(y_1-y_3)(y_2-y_4)}
$$
The $e(\tau)\to 0$ limit is the weakly-coupled $Spin(8)$ gauge theory; $e(\tau)\to \infty$ is the weakly-coupled $SU(2)$ gauge theory and $e(\tau)\to 1$ yields the weakly-coupled $G_2$ gauge theory.

The other cases are equally-easy to write down. It would be interesting to compare these results with the Seiberg-Witten curves obtained in \cite{Aganagic:1997wp, Terashima:1998fx}.

\section{{$Spin(7)$ Gauge Theory}}\label{Spin7}

$Spin(7)$, with $n$ hypermultiplets in the $8$ and $(5-n)$ in the $7$, also has vanishing $\beta$-function. Perhaps with the addition of some free hypermultiplets, we can realize the cases $n=2,3,4,5$ in the $D_4$ theory.

\subsection{{$2(8)+3(7)$}}\label{Spin7-23}

This theory (with the addition of two free hypermultiplets) was one of the examples discussed in \S\ref{Sp5}. The theory has two strong-coupling points.

\begin{itemize}%
\item One is a $G_2$ gauge theory, with two hypermultiplets in the $7$, coupled to the ${(E_7)}_8$ SCFT.
\item The other is an $SU(2)$ gauge theory coupled to the ${Sp(4)}_8\times {Sp(2)}_7$ SCFT.

\end{itemize}
\subsection{{$3(8)+2(7)$}}\label{Spin7-32}

This theory (with the addition of two free hypermultiplets) was discussed in \S\ref{Sp4Sp2}. The S-dual theory is an $Sp(2)$ gauge theory with 5 half-hypermultiplets in the $4$, coupled to the ${Sp(5)}_7$ SCFT.

\subsection{{$4(8)+1(7)$}}\label{Spin7-41}

This theory (with the addition of one free hypermultiplet) was also discussed in \S\ref{Sp4Sp2}. The S-dual theory is an $SU(2)$ gauge theory with a half-hypermultiplet in the $2$, coupled to the ${Sp(4)}_8\times {Sp(2)}_7$ SCFT.

\subsection{{$5(8)$}}\label{Spin7-50}

This theory has three degeneration limits, two of which

\begin{displaymath}
\begin{gathered}
\begin{matrix} \includegraphics[width=279pt]{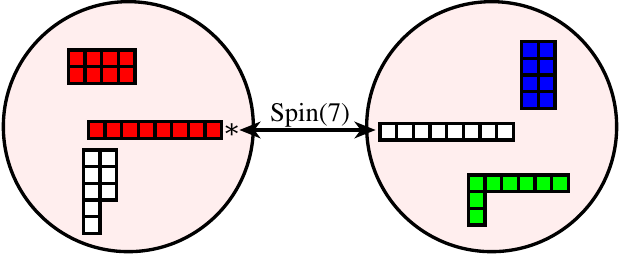}\end{matrix}\\
\begin{matrix} \includegraphics[width=279pt]{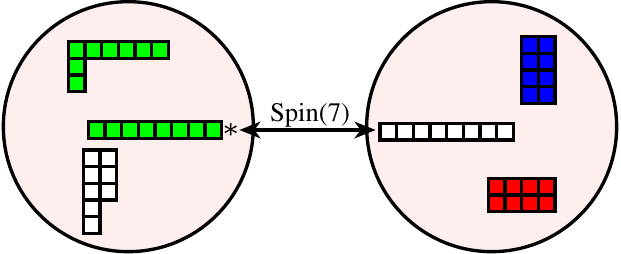}\end{matrix}
\end{gathered}
\end{displaymath}
are $Spin(7)$ gauge theories with matter in the $5(8)$. The fixture on the left contributes $2(8)$; the fixture on the right contributes $3(8)$.

The other degeneration,

\begin{displaymath}
\begin{matrix} \includegraphics[width=279pt]{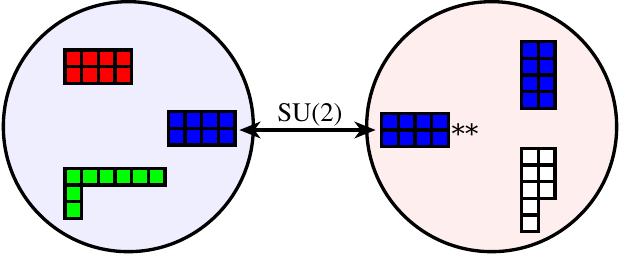}\end{matrix}
\end{displaymath}
is an $SU(2)$ gauge theory coupled to the $Sp(6)_8$ SCFT (the fixture on the right is empty).

\section{{Other Interesting Examples}}\label{Other}

\subsection{Fun with interacting SCFTs}\label{fun}
Let us take the ${Sp(2)}^3_8\times {SU(2)}_7$ SCFT and gauge an ${SU(2)}_8$ subgroup (the fixture on the right is empty):

\begin{displaymath}
\begin{aligned}
\begin{matrix} \includegraphics[width=279pt]{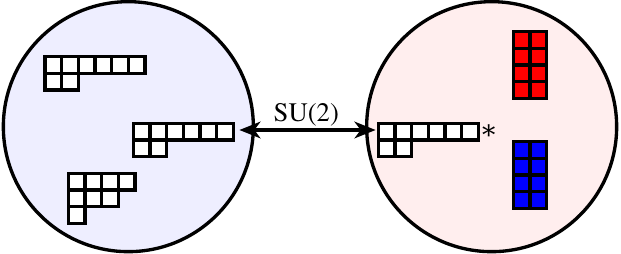}\end{matrix}
\end{aligned}
\end{displaymath}
The S-dual theory is

\begin{displaymath}
\begin{aligned}
\begin{matrix} \includegraphics[width=279pt]{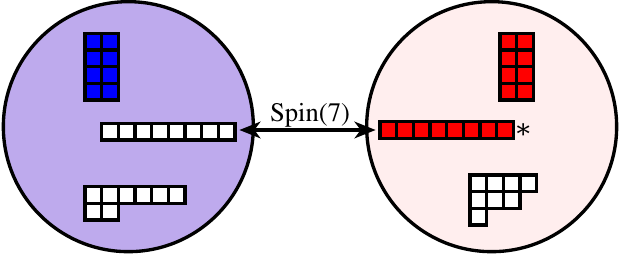}\end{matrix}
\end{aligned}
\end{displaymath}
The fixture on the right contributes hypermultiplets in the $7+8$. The fixture on the left is the ${(E_7)}_8$ SCFT with matter in the $8_c$ of $Spin(8)$. Under the given embedding of $Spin(7)$, this matter transforms as an additional $8$. So the matter contributes an ${Sp(2)}_8\times {SU(2)}_7$ to the global symmetry group of the theory. The rest, ${Sp(2)}_8\times {SU(2)}_8$, is the centralizer of $Spin(7)\subset E_7$.

As another example of our methods, let us consider various gaugings of the ${Sp(2)}^2_8\times {SU(2)}_8^4$ SCFT. We can gauge an $Sp(2)$ subgroup,

\begin{displaymath}
\begin{aligned}
\begin{matrix} \includegraphics[width=279pt]{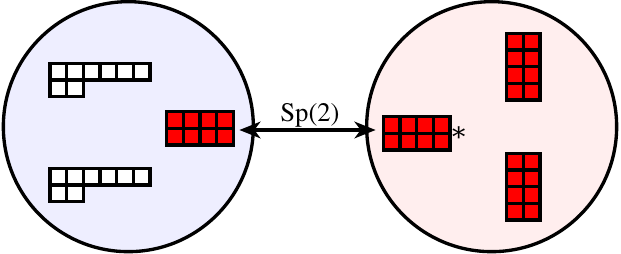}\end{matrix}
\end{aligned}
\end{displaymath}
where the fixture on the right provides two hypermultiplets in the fundamental of $Sp(2)$. The S-dual theory,

\begin{displaymath}
\begin{aligned}
\begin{matrix} \includegraphics[width=279pt]{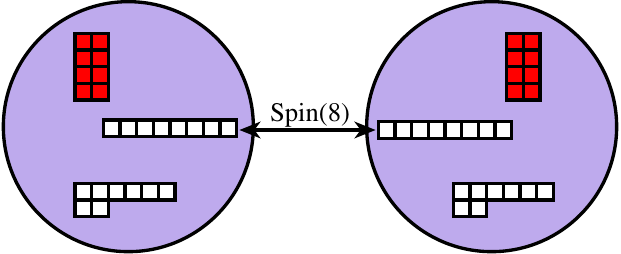}\end{matrix}
\end{aligned}
\end{displaymath}
is a $Spin(8)$ gauge theory, with matter in the $2(8_s)$, coupled to two copies of the ${(E_7)}_8$ SCFT.

Instead, we can gauge an $SU(2)$ subgroup

\begin{displaymath}
\begin{aligned}
\begin{matrix} \includegraphics[width=279pt]{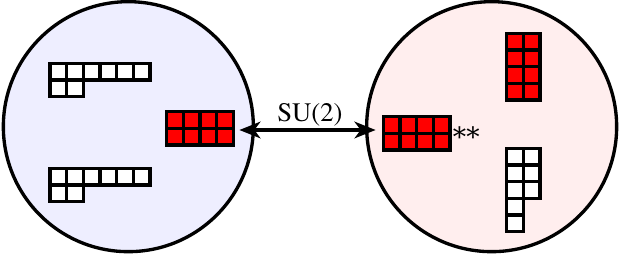}\end{matrix}
\end{aligned}
\end{displaymath}
where the fixture on the right is empty. The S-dual

\begin{displaymath}
\begin{aligned}
\begin{matrix} \includegraphics[width=279pt]{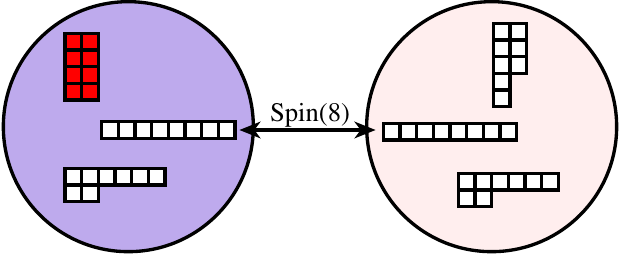}\end{matrix}
\end{aligned}
\end{displaymath}
is a $Spin(8)$ gauge theory, with matter in the $2(8_s)+8_c+8_v$, coupled to one copy of the ${(E_7)}_8$ SCFT.

A different $SU(2)$ gauging of the ${Sp(2)}^2_8\times {SU(2)}^4_8$ SCFT

\begin{displaymath}
\begin{aligned}
\begin{matrix} \includegraphics[width=279pt]{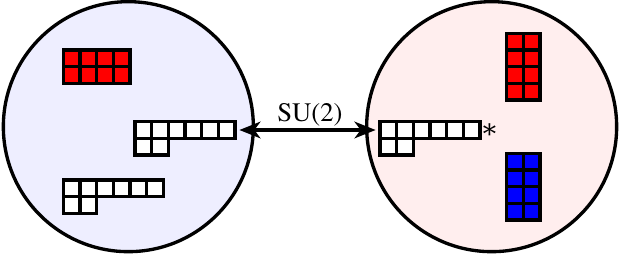}\end{matrix}
\end{aligned}
\end{displaymath}
has two distinct strong-coupling points. One,

\begin{displaymath}
\begin{aligned}
\begin{matrix} \includegraphics[width=279pt]{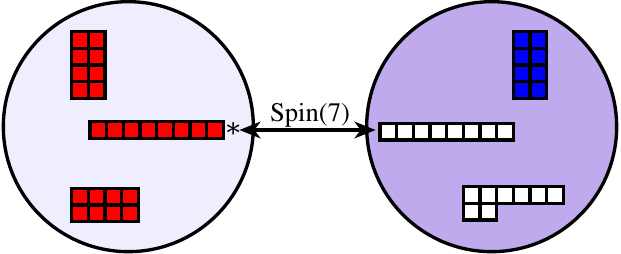}\end{matrix}
\end{aligned}
\end{displaymath}
is a $Spin(7)$ gauge theory, with matter in the $8$, coupled to two copies of the ${(E_7)}_8$ SCFT. The other,

\begin{displaymath}
\begin{aligned}
\begin{matrix} \includegraphics[width=279pt]{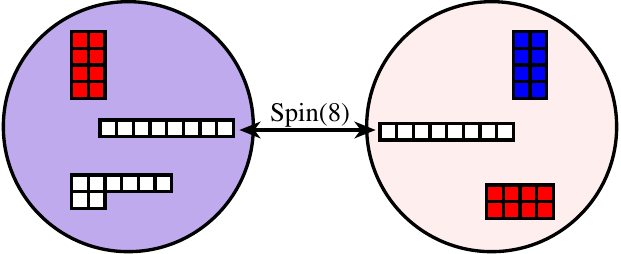}\end{matrix}
\end{aligned}
\end{displaymath}
is a $Spin(8)$ gauge theory, with matter in the $2(8_s)+2(8_c)$, coupled to a single copy of the ${(E_7)}_8$ SCFT.

\subsection{{$D_5$ example: $Spin(10)$ gauge theory}}\label{Spin10}

To further illustrate our methods, let us study \emph{one} example from the $D_5$ theory, involving a $Spin(10)$ gauge theory with matter in the $3(16)+2(10)$.

Start with the 4-punctured sphere

\begin{displaymath}
\begin{matrix} \includegraphics[width=287pt]{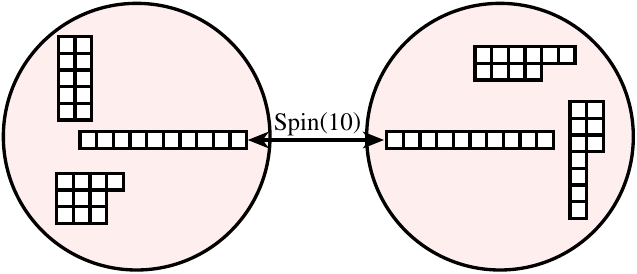}\end{matrix}
\end{displaymath}
This is a Spin(10) Lagrangian field theory with matter in the $3(16)+2(10)$ representation. The left fixture provides 32 free hypermultiplets in the $(16,2)$ of $Spin(10)\times SU(2)$, and the right fixture, 36 free hypermultiplets in the $(16,1)+\tfrac{1}{2}(10,4)$ of $Spin(10)\times Sp(2)$.

The global symmetry group of the theory is, thus,

\begin{displaymath}
G_{\text{global}}=SU(3)_{32} \times Sp(2)_{10} \times U(1),
\end{displaymath}
This theory has two distinct strong coupling cusp points. One appears in the degeneration

\begin{displaymath}
\begin{matrix} \includegraphics[width=287pt]{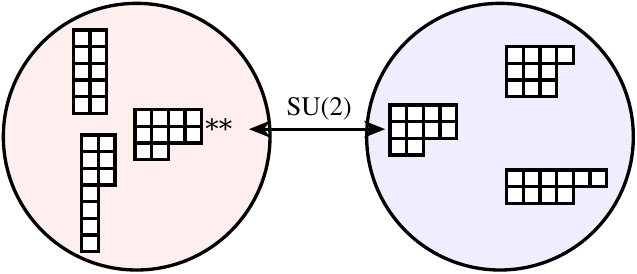}\end{matrix}
\end{displaymath}
Here the left fixture is empty. The $\begin{matrix} \includegraphics[width=45pt]{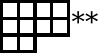}\end{matrix}$ irregular puncture has pole structure $\{1,5,7,10;6\}$, and imposes the constraint $c^{(8)}_{10}={(c^{(4)}_5)}^2$. The right fixture is an interacting SCFT with graded Coulomb branch dimension $d=(0,0,1,1,1,0,1)$ and global symmetry group

\begin{displaymath}
G_{\text{SCFT}}=Sp(2)_{10}\times SU(3)_{32}\times SU(2)_{8}\times U(1),
\end{displaymath}
and we gauge the $SU(2)_8$ subgroup.

The second strong coupling point appears in the remaining degeneration,

\begin{displaymath}
\begin{matrix} \includegraphics[width=287pt]{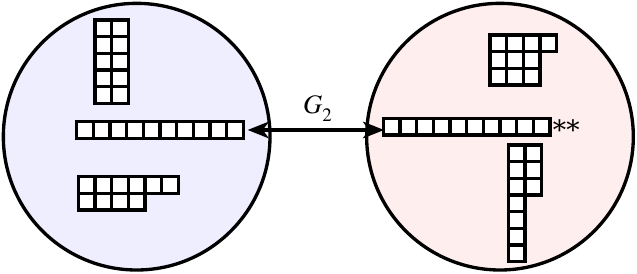}\end{matrix}
\end{displaymath}
Here the fixture on the left is an SCFT with graded Coulomb branch dimension $d=(0,0,1,1,0,0,1)$ and global symmetry group

\begin{displaymath}
G_{\text{SCFT}} = {(E_6)}_{16} \times {Sp(2)}_{10} \times U(1),
\end{displaymath}
and the fixture on the right is empty. The $\begin{matrix} \includegraphics[width=92pt]{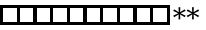}\end{matrix}$ irregular puncture has pole structure $\{1,4,5,8;5\}$. Under the decomposition ${(E_6)}_k\supset {(G_2)}_k \times {SU(3)}_{2k}$, we gauge a $(G_2)_{16} \subset (E_6)_{16}$.

\section*{Acknowledgements}
The research of the authors is based upon work supported by the National Science Foundation under Grant No. PHY-0969020. The work of J.~D.~was also supported by the United States-Israel Binational Science Foundation under Grant \#2006157. J.~D.~would like to thank the Erwin Schr\"odinger Institute, in Vienna, for hospitality while this manuscript was completed. We have benefited tremendously from conversations with Andrew Neitzke and Yuji Tachikawa, as well as from some useful remarks of David Ben-Zvi.

In the first version of this paper, for seven entries in the table of interacting fixtures of the $D_4$ theory, we had misidentified the global symmetry group of the SCFT. We would like to thank Simone Giacomelli and Yuji Tachikawa for pointing out the enhanced global symmetry for some of those entries, which led us to uncover the rest. We would also like to thank Andy Trimm for checking some of those calculations.
\vfill\eject
\appendix

\section{{Appendix: Nilpotent orbits in $\mathfrak{so}(2N)$}}\label{appendix}

Here we lay out our conventions for nilpotent orbits in $\mathfrak{so}(2N)$. For more details, see \cite{Collingwood1993}. We take $\mathfrak{so}(2N)$ to consist of block matrices of the form

\begin{equation}
\begin{pmatrix}A& B\\ C& -A^t\end{pmatrix}
\label{so2Nalg}\end{equation}
where $A,B,C$ are $N\times N$ matrices and $B^t=-B$, $C^t=-C$. Nilpotent orbits are in 1-1 correspondence with embeddings $\rho: \mathfrak{sl}(2)\hookrightarrow \mathfrak{so}(2N)$, up to conjugation. Here, $\mathfrak{sl}(2)$ is generated by $\{H,X,Y\}$ satisfying

\begin{equation}
[H,X] = 2X,\quad [H,Y]=-2Y,\quad [X,Y]=H
\label{sl2alg}\end{equation}
and we take $\rho(X)$ (which we will, henceforth, simply denote by $X$) as our representative element of the nilpotent orbit.

As noted in the text, a nilpotent orbit, in $\mathfrak{so}(2N)$, is specified by a D-partition of $2N$. Here, we will give our convention for assigning a triple of matrices of the form \eqref{so2Nalg}, satisfying \eqref{sl2alg}, to such a partition.

Let $e_1,e_2,\dots e_n$ be the standard basis for $\mathbb{C}^N$. Let $E_{i,j}$ be the $2N\times 2N$ matrix with a $1$ in the ${(i,j)}^{\text{th}}$ position and zeroes everywhere else. To the root, $e_i-e_j$, assign the matrix, of the form \eqref{so2Nalg},

\begin{displaymath}
X^-_{i,j} = E_{i,j}-E_{j+N,i+N}
\end{displaymath}
To the root $e_i+e_j$ (for $i\lt j)$, assign

\begin{displaymath}
X^+_{i,j} = E_{i,j+N}- E_{j,i+N}, \quad i\lt j
\end{displaymath}
Also, let

\begin{displaymath}
H_i = E_{i,i}-E_{i+N,i+N}
\end{displaymath}
\begin{itemize}%
\item Take the D-partition, $[r_1,r_2,\dots]$, and divide it into pairs of the form $[r,r]$ and $[2s+1,2t+1]$ ($s\gt t$). This is not quite unique: the $D_6$ partition, $[3,3,2,2,1,1]$ can be divided into $[3,3],\, [2,2],\, [1,1]$ or into $[2,2],\,[3,1],\, [3,1]$. Different choices will result in different representatives of the same nilpotent orbit.

\item To each pair of the form $[r,r]$, assign a block of $r$ consecutive basis vectors of $\mathbb{C}^N$. We will denote those by $(e_1,e_2,\dots,e_r)$, but they might be, say, $(e_{17},e_{18},\dots,e_{16+r})$. To each pair of the form $[2s+1,2t+1]$, assign a block of $s+t$ consecutive basis vectors of $\mathbb{C}^N$. The blocks, thus assigned, must be non-overlapping, and will exhaust $e_1,\dots,e_N$.

\item For each pair of the form $[r,r]$, let

\begin{displaymath}
\begin{aligned}
H&= \sum_{k=1}^r (r+1-2k) H_k\\
X&=\sum_{k=1}^{r-1} \sqrt{k(r-k)} X^-_{k,k+1}\\
Y&= X^t
\end{aligned}
\end{displaymath}

\item For pairs of the form $[2s+1,2t+1]$, the general formula can be found in \cite{Collingwood1993}. We will need just the first few, for small values of $t$.

\begin{itemize}%
\item For pairs of the form $[2s+1,1]$, let

\begin{displaymath}
\begin{aligned}
H&=\sum_{k=1}^s 2(s+1-k)H_k\\
X&=\sum_{k=1}^{s-1}\sqrt{k(2s+1-k)}X^-_{k,k+1}+\sqrt{s(s+1)/2}\left(X^-_{s,s+1}+X^+_{s,s+1}\right)\\
Y&=X^t
\end{aligned}
\end{displaymath}

\item For pairs of the form $[2s+1,3]$, let

\begin{displaymath}
\begin{aligned}
H&=\sum_{k=1}^s 2(s+1-k)H_k\, + 2 H_{s+1}\\
X&= \sum_{k=1}^{s-2} \sqrt{k(2s+1-k)} X^-_{k,k+1}
  + \sqrt{(s-1)(s+2)}X^-_{s-1,s}\\
 &\quad+ \sqrt{s(s+1)/2}\left(X^-_{s,s+2}+X^+_{s,s+2}\right)
            + \left(X^-_{s+1,s+2}-X^+_{s+1,s+2}\right)\\
Y&=X^t
\end{aligned}
\end{displaymath}

\end{itemize}

\item Add up the contributions to $H, X, Y$ from each pair. The resulting triple, $\{H,X,Y\}$, will be our embedding of $\mathfrak{sl}(2)$ and $X$ will be our representative of the nilpotent orbit, corresponding to this partition.

\end{itemize}
The one exception to this rule has to do with ``{}very even''{} partitions and our red/blue\footnote{Our ``red'' and ``blue'' Hitchin D-partitions correspond, respectively, to the partitions with labels ``I'' and ``II'' in Recipe 5.2.6 of \cite{Collingwood1993}.} nilpotent orbits.

\begin{itemize}%
\item For the red orbit, follow the prescription above.
\item For the blue orbit, replace every instance of $X^\mp_{i,N}$ with $X^\pm_{i,N}$ and replace every instance of $H_N$ with $-H_N$. This has the effect of exchanging the roles of the two irreducible spinor representations and flips the sign of the Pfaffian, $\tilde{\phi}(y)\to - \tilde{\phi}(y)$.

\end{itemize}

\vfill\eject
\bibliographystyle{utphys}
\bibliography{gaiotto}

\end{document}